\documentclass[american,]{article}
\usepackage{times}
\usepackage{amssymb,amsmath,amsthm}
\usepackage{ifxetex,ifluatex}
\usepackage{fixltx2e} 
\ifnum 0\ifxetex 1\fi\ifluatex 1\fi=0 
  \usepackage[T1]{fontenc}
  \usepackage[utf8]{inputenc}
\else 
  \ifxetex
    \usepackage{mathspec}
  \else
    \usepackage{fontspec}
  \fi
  \defaultfontfeatures{Ligatures=TeX,Scale=MatchLowercase}
\fi
\IfFileExists{upquote.sty}{\usepackage{upquote}}{}
\IfFileExists{microtype.sty}{%
\usepackage{microtype}
\UseMicrotypeSet[protrusion]{basicmath} 
}{}
\usepackage[margin=1in]{geometry}
\usepackage{hyperref}
\hypersetup{unicode=true,
            pdftitle={Rank-normalization, folding, and localization: An improved R-hat for assessing convergence of MCMC},
            pdfauthor={Aki Vehtari, Andrew Gelman, Daniel Simpson, Bob Carpenter, Paul-Christian Bürkner},
            pdfkeywords={keywords},
            pdfborder={0 0 0},
            breaklinks=true,
            colorlinks=true,
            linkcolor=black,
            citecolor=black,
            filecolor=black,
            urlcolor=black,
          }
\ifnum 0\ifxetex 1\fi\ifluatex 1\fi=0 
  \usepackage[shorthands=off,main=american]{babel}
\else
  \usepackage{polyglossia}
  \setmainlanguage[variant=american]{english}
\fi
\usepackage{natbib}
\usepackage{placeins}
\usepackage{graphicx,grffile,subcaption}
\makeatletter
\def\maxwidth{\ifdim\Gin@nat@width>\linewidth\linewidth\else\Gin@nat@width\fi}
\def\maxheight{\ifdim\Gin@nat@height>\textheight\textheight\else\Gin@nat@height\fi}
\makeatother
\setkeys{Gin}{width=\maxwidth,height=\maxheight,keepaspectratio}
\IfFileExists{parskip.sty}{%
\usepackage{parskip}
}{
\setlength{\parindent}{0pt}
\setlength{\parskip}{6pt plus 2pt minus 1pt}
}
\providecommand{\tightlist}{%
  \setlength{\itemsep}{0pt}\setlength{\parskip}{0pt}}
\ifx\paragraph\undefined\else
\let\oldparagraph\paragraph
\renewcommand{\paragraph}[1]{\oldparagraph{#1}\mbox{}}
\fi
\ifx\subparagraph\undefined\else
\let\oldsubparagraph\subparagraph
\renewcommand{\subparagraph}[1]{\oldsubparagraph{#1}\mbox{}}
\fi

\let\rmarkdownfootnote\footnote%
\def\footnote{\protect\rmarkdownfootnote}

\usepackage{titling}

\usepackage{booktabs}
\usepackage{longtable}
\usepackage{array}
\usepackage{multirow}
\usepackage[table]{xcolor}
\usepackage{wrapfig}
\usepackage{float}
\usepackage{colortbl}
\usepackage{pdflscape}
\usepackage{tabu}
\usepackage{threeparttable}
\usepackage{threeparttablex}
\usepackage[normalem]{ulem}
\usepackage{makecell}

\usepackage{mathtools}
\usepackage[utf8]{inputenc}
\usepackage[T1]{fontenc}
\usepackage{textcomp}
\usepackage{graphicx,pdflscape}
\usepackage{geometry}
\usepackage{amsmath}
\usepackage{float}
\usepackage{supertabular}
\usepackage{booktabs,caption}
\usepackage{tcolorbox}
\usepackage{paralist}
\usepackage{multicol}
\setcitestyle{round}

\DeclareMathOperator{\N}{Normal}
\DeclareMathOperator{\Gam}{Gamma}
\DeclareMathOperator{\Beta}{Beta}
\DeclareMathOperator{\Var}{Var}
\DeclareMathOperator{\var}{var}
\DeclareMathOperator{\E}{E}
\DeclareMathOperator{\I}{I}
\newcommand{\Rhat}{$\widehat{R}$}
\newcommand{\sRhat}{split-$\widehat{R}$}
\theoremstyle{definition}

\title{Rank-normalization, folding, and localization:\\
  An improved $\widehat{R}$ for assessing convergence of MCMC\footnote{To appear in Bayesian Analysis. We thank Ben Bales, Ian Langmore, the editor, and anonymous reviewers for useful comments. We also thank Academy of Finland, the U.S. Office of Naval Research, National Science Foundation, Institute for Education Sciences, the Natural Science and Engineering Research Council of Canada, Finnish Center for Artificial Intelligence, and Technology Industries of Finland Centennial Foundation for partial support of this research.  All computer code and an 
even larger variety of numerical experiments are available in the online 
appendix at \url{https://avehtari.github.io/rhat_ess/rhat_ess.html}.}\vspace{.1in}}
    \pretitle{\vspace{\droptitle}\centering\huge}
  \posttitle{\par}
    \author{Aki Vehtari, Andrew Gelman, Daniel Simpson, Bob Carpenter, Paul-Christian Bürkner}
    
    \author{
Aki Vehtari\footnote{Department of Computer Science, Aalto University, Finland.},
   Andrew Gelman\footnote{Department of Statistics, Columbia University, New York.},
 Daniel Simpson\footnote{Department of Statistical Sciences, University of Toronto, Canada.},
 Bob Carpenter\footnote{Center for Computational Mathematics, Flatiron Institute, New York.},
and Paul-Christian B\"{u}rkner$^\dagger$
}
    
    \preauthor{\centering\large\emph}
  \postauthor{\par}
     \date{}

\begin{document}
\maketitle
\begin{abstract}
  Markov chain Monte Carlo is a key computational tool in Bayesian 
  statistics, but it can be challenging to monitor the convergence of an iterative stochastic algorithm.
In this paper we show that the convergence diagnostic \Rhat\ 
of \citet{Gelman+Rubin:1992} has serious flaws. Traditional \Rhat\ will fail to correctly diagnose
convergence failures when the chain has a heavy tail or when the variance varies across 
the chains. In this paper we propose an alternative rank-based diagnostic that fixes these 
problems. We also introduce
  a collection of quantile-based local efficiency
  measures, along with a practical approach for computing Monte Carlo error
  estimates for quantiles. We suggest that common trace plots should
  be replaced with rank plots from multiple chains. Finally, we give
  recommendations for how these methods should be used
  in practice.
\end{abstract}

\hypertarget{introduction}{%
\section{Introduction}\label{introduction}}

Markov chain Monte Carlo (MCMC) methods are important in computational statistics, especially 
in Bayesian applications where the goal is to represent
posterior inference using a sample of posterior draws. While MCMC, 
as well as more general iterative
simulation algorithms, can usually be proven to converge
to the target distribution as the number of draws approaches infinity,
there are rarely strong guarantees about their behavior after finite time. Indeed, decades of experience tell us that
the finite sample behavior of these algorithms can be almost arbitrarily bad.

\subsection{Monitoring convergence using multiple chains}

In an attempt to assuage concerns of poor convergence, we typically run multiple 
independent chains  to see if the obtained 
distribution is similar across chains.  We can also visually inspect
the sample paths of the chains via trace plots as well as study summary 
statistics such as the empirical autocorrelation function. 

\begin{figure}
\center
  \begin{subfigure}[b]{0.37\textwidth}
    \includegraphics[width=\textwidth]{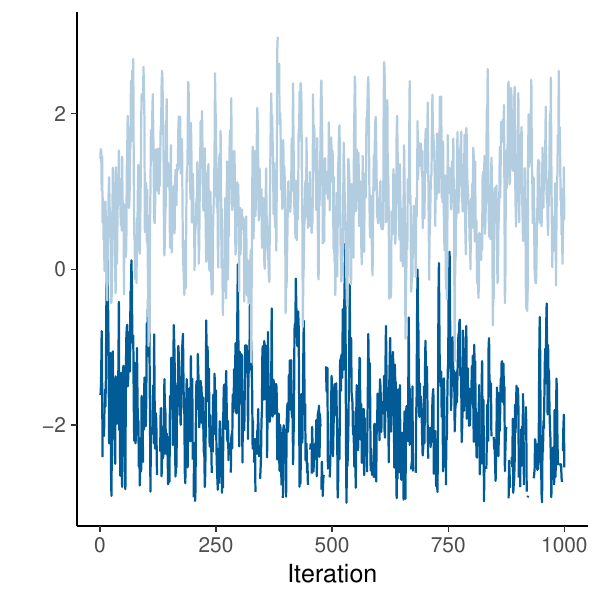}
  \end{subfigure}
  \hspace{8mm}
\begin{subfigure}[b]{0.37\textwidth}
    \includegraphics[width=\textwidth]{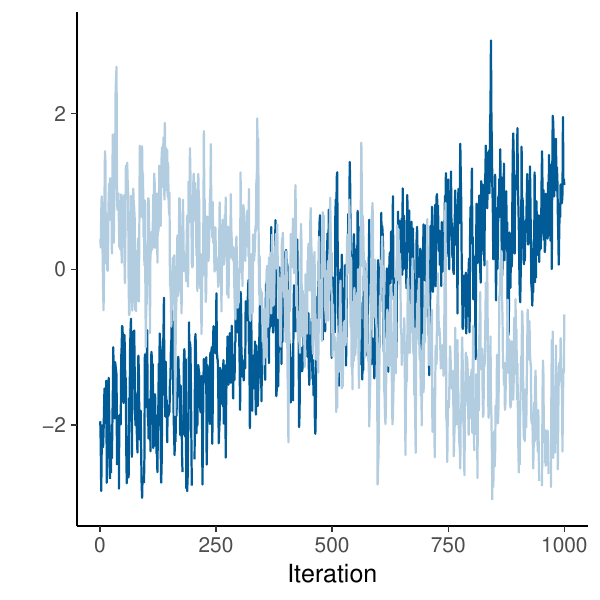}
\end{subfigure}
\caption{Examples of two challenges in assessing convergence of iterative
simulations. (a) In the left plot, either sequence alone looks stable, but the
juxtaposition makes it clear that they have not converged to a common
distribution. (b) In the right plot, the two sequences happen to cover a common
distribution but neither sequence appears stationary. These graphs demonstrate
the need to use between-sequence and also within-sequence information when
assessing convergence. Adapted from \citet{BDA3}.}
\label{converge.challenge}
\end{figure}

Running multiple chains is critical to any MCMC convergence diagnostic. Figure
\ref{converge.challenge} illustrates two ways in which sequences of
iterative simulations can fail to converge.  In the first example, two chains
are in different parts of the target distribution; in the second
example, the chains move but have not attained stationarity. Slow mixing can arise with multimodal target distributions or when a chain is
stuck in a region of high curvature with a step size too large to make an
acceptable proposal for the next step. The two examples in Figure \ref{converge.challenge}  make it clear that 
any method for assessing mixing and effective sample size should use information
between and within chains.

As we are often fitting models with large
numbers of parameters, it is not realistic to expect to make and interpret
trace plots such as in Figure \ref{converge.challenge} for all
quantities of interest. Hence we need numerical summaries that can flag
potential problems. 

Of the various convergence diagnostics \citep[see reviews by][]{Cowles+Carlin:1996,Mengersen+etal:1999,Robert+Casella:2004}, 
probably the most widely used  
is the potential scale reduction factor \Rhat\
\citep{Gelman+Rubin:1992, Brooks+Gelman:1998}.
It is recommended as the primary convergence diagnostic in widely applied
software packages for MCMC sampling such as Stan \citep{Stan:JSS:2017}, 
JAGS \citep{plummer2003jags}, WinBUGS \citep{WinBUGS:2000}, OpenBUGS \citep{BUGSproject:2009}, PyMC3 \citep{pymc3}, 
and NIMBLE \citep{nimble}, which together are estimated to have hundreds of thousands of users. 
\Rhat\ is computed for each scalar quantity of interest, as the standard deviation of that quantity from all the chains included together, divided by the root mean square of the separate within-chain standard deviations.
The idea is that if a set of simulations have not mixed well, the variance of
all the chains mixed together should be higher than the variance of individual chains.
More recently, \cite{BDA3} introduced \sRhat\ which also compares 
the first half of each chain to the second
half, to try to detect lack of convergence within each chain.  In this
paper when we refer to \Rhat\ we are always speaking of the \sRhat\ variant.

Convergence diagnostics are most effective when computed using multiple chains initialized at a 
diverse set of starting points. This reduces the chance that we falsely diagnose
mixing when beginning at a different point would lead to a 
qualitatively different posterior.

In the context of Markov chain Monte Carlo, one can interpret \Rhat\ 
with diverse seeding as an operationalization of the qualitative statement 
that, after warmup, convergence of the Markov chain should be relatively insensitive to the starting 
point, at least within a 
reasonable part of the parameter space. This is the closest we can come to 
verifying empirically that the Markov chain is geometrically ergodic, which is a critical 
property if we want  a central limit theorem to hold for approximate
posterior expectations. Without this, we have no control over the large
deviation behavior of the estimates and the constructed Markov chains may
be useless for practical purposes.

\subsection{Example where traditional \Rhat\ fails}

Unfortunately, \Rhat\ can fail to diagnose poor mixing, which can be a problem when it is used as a default rule. The following example shows how
failure can occur.

\begin{figure}
\centering
\includegraphics[width=0.95\textwidth]{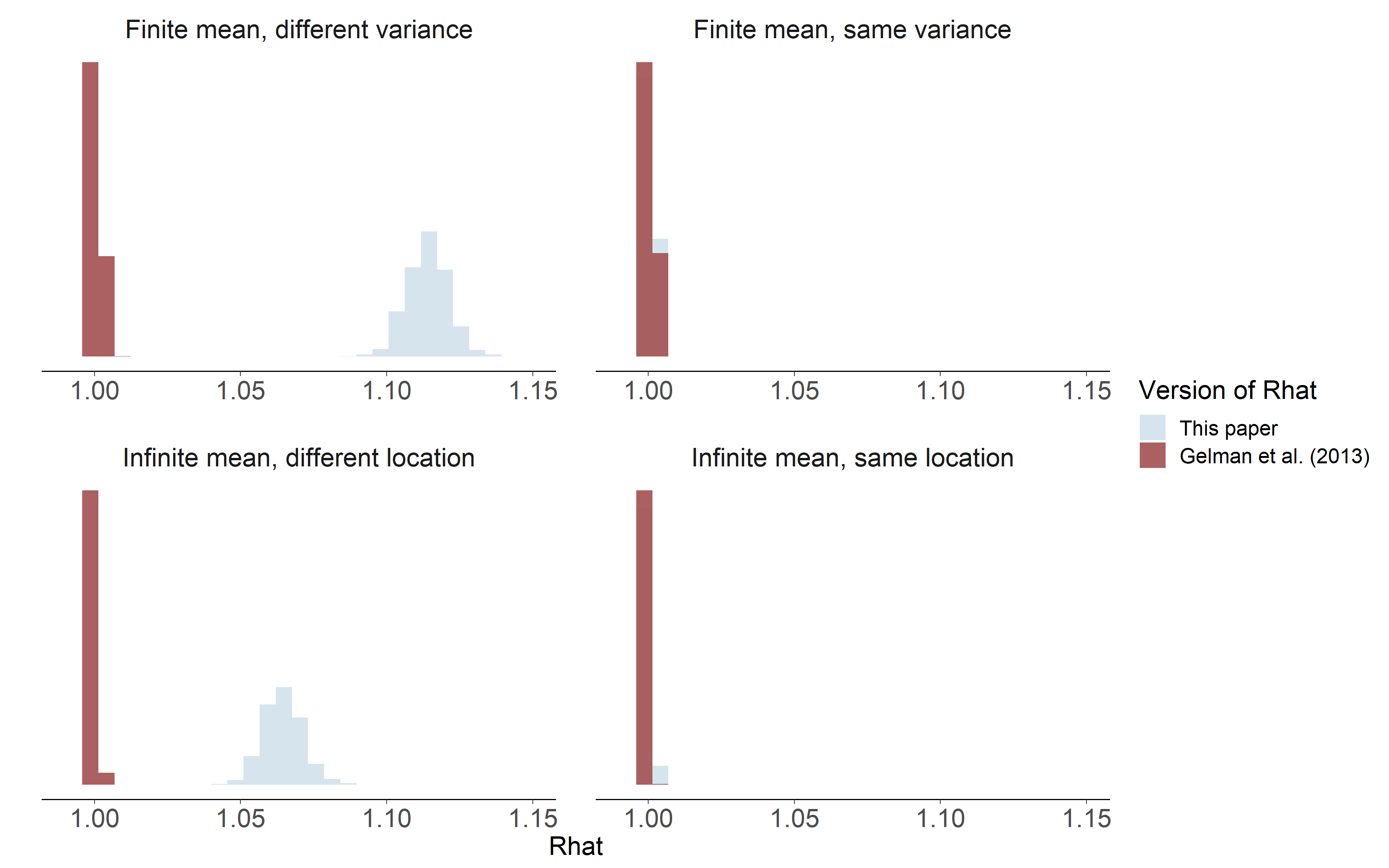}
\caption{An example showing problems undetected by
traditional \Rhat\ . Each plot shows
histograms of \Rhat\ values over $1000$ replications of four chains, each
with a thousand draws. In the left column, one of these four chains was 
incorrect. In the top left plot, we set one of the four chains to have a variance lower than the others.
In the bottom left plot, we took one of the four chains and shifted it.
In both cases, the traditional \Rhat\ estimate does not detect the poor
behavior, while the new value does. In the right column, all the chains are
simulated with the same distribution. The chains used for the top row plots target 
a normal distribution, while the chains used for the bottom row plots target
a Cauchy distribution. \label{fig:simple_example}}
\end{figure}

The red histograms in Figure~\ref{fig:simple_example} show the distribution of \Rhat\  (that is, \sRhat\ from \cite{BDA3}) in four different scenarios.  (Ignore the light blue histograms for now; they show the results using an improved diagnostic that we shall discuss later in this paper.)  In all four scenarios, traditional \Rhat\ is well under 1.1 under all simulations, thus not detecting any convergence problems---but in fact the two scenarios on the left have been constructed so that they are far from mixed.  These are problems that are not detected by traditional \Rhat.

In each of the four scenarios in  Figure~\ref{fig:simple_example}, we run four chains for $1000$ iterations each and then replicate the entire simulation 1000 times. The top row of the figure shows results for independent AR(1) processes with autoregressive parameter $\rho=0.3$. The top left graph shows the distribution of \Rhat\ when one of the four chains is manually transformed to only have $1/3$ of the variance 
compared to the other three chains (see Appendix~A for more details). This corresponds to a scenario where one chain fails to correctly explore the tails of the target distribution and one would hope could be identified as non-convergent. The \sRhat\ statistic defined in \citet{BDA3} does not detect the poor mixing, while the new variant of \sRhat\ defined later in this paper does. The top-right figure shows the same scenario but with all the chains having the same variance, and now both \Rhat\ values correctly identify that mixing occurs.

The second row of Figure~\ref{fig:simple_example} shows the behavior of \Rhat\ when the target distribution has infinite variance. In this case the chains were constructed as a ratio of stationary AR(1) processes with  $\rho=0.3$, and the distribution of the ratio is Cauchy.  All of the simulated chains have  unit scale, but in the lower-left figure, we have manually shifted one of the four chains two units to the right. This corresponds to a scenario where one chain provides a biased estimate of the target distribution. The \citet{BDA3} version of \Rhat\ would catch this behavior if the chain had finite variance, but in this case the infinite variance destroys its effectiveness---traditional  \Rhat\ and \sRhat\ are defined based on second-moment statistics---and it inappropriately returns a value very close to 1.

This example identified two problems with traditional \Rhat : 
\begin{enumerate}
\item If the chains have different variances but the same mean parameters, traditional $\widehat{R} \approx 1$.
\item If the chains have infinite variance, traditional $\widehat{R} \approx 1$ even if one of the chains has a different location parameter to the others. This can also lead to numerical instability for thick-tailed distributions even when the variance is technically finite. It's typically hard to assess empirically
if a chain has large but finite variance or infinite variance.
\end{enumerate}

A related problem is that \Rhat\ is
typically computed only for the posterior mean. While this provides an estimate 
for the convergence in the bulk of the distribution, it says little about the 
convergence in the tails, which is a concern for posterior 
interval estimates as well as for inferences about rare events.
 


\section{Recommendations for practice}

The traditional \Rhat\ statistic is general, easy to compute, and can
catch many problems of poor convergence, but the discussion above
reveals some scenarios where it fails. The present paper proposes
improvements that overcome these problems.
In addition, as the convergence
of the Markov chain needs not be uniform across the parameter space, we
propose a localized version of effective sample size
that allows us to assess better the behavior of localized 
functionals and quantiles of the chain.
Finally, we propose three new methods to visualize the 
convergence of an iterative algorithm that are more informative than standard 
trace plots.

In this section we lay out practical recommendations for using the tools 
developed in this paper. In the interest of specificity, we have 
provided numerical targets for both \Rhat\ and effective sample size (ESS),
which are useful as first level checks when analyzing reliability of inference for
many quantities. However, these values should be adapted as necessary for the given
application, and ultimately domain expertise should be used to check that Monte Carlo
standard errors (MCSE) for all quantities of interest are small enough.

In Section \ref{improving-convergence-diagnostics}, we propose modifications to  
\Rhat\ based on rank-normalizing and folding the posterior draws, only using the sample if  $\widehat{R} < 1.01$. 
This threshold is much tighter than the one recommended by 
\citet{Gelman+Rubin:1992}, reflecting lessons learnt over more than 25 years of use, as well as
the simulation results in Appendix~A.  \citet{Gelman+Rubin:1992}
derived \Rhat\ under the assumption that, as simulations went forward,
the within-chain variance would gradually increase while the
between-chain variance decreased, stabilizing when their ratio was 1.
The potential scale reduction factor represented the factor by which
the between-chain variation might decline under future simulations,
and a potential scale reduction factor of 1.1 implied that there was
little to be gained in inferential precision by running the chains
longer.  However, as discussed by \citet{Brooks+Gelman:1998}, the
dynamics of MCMC are such that the between-chain variance can decrease
before it increases, if the initial part of the simulation pulls all
the chains to the center of the distribution, only for them to be
redispersed with further simulation. As a result, \Rhat\ cannot in
general be interpreted as a potential scale reduction factor, and in
practice and in simulations we have found that \Rhat\ can dip below
1.1 well before convergence in some examples (a point also raised by
\citet{vats2018revisiting}), and we have found
this to be much more rare when using the 1.01 threshold.

In addition, we recommend running at least 
four chains by default.
Multiple chains are more likely to reveal multimodality and poor adaptation or mixing:  we see examples for complex,
misspecified or non-identifiable models in the Stan discussion forum all
the time. Furthermore, most computers are able to run chains in
parallel, giving multiple chains with no increase in computation time.
Here we do not consider massive parallelization such as running 1000
chains or more; further research is needed in considering how to use
such simulations most efficiently in such computational environments (see, for instance, the 
method discussed in \citet{jacob2017unbiased}).

Roughly speaking, the effective sample size of a quantity of interest captures how many
independent draws contain the same amount of information as the dependent 
sample obtained by the MCMC algorithm. The higher the ESS the better.
When there might be difficulties with mixing, it is important to use between-chain as well as within-chain
information in computing the ESS. A common example arises in hierarchical models with funnel-shaped posteriors, where MCMC algorithms can struggle to simultaneously adapt to a ``narrow'' region of 
high density and low volume, and a ``wide'' region of low density and high volume. In such a case, differences in step-size adaptation can
lead to chains that have different behavior in the neighborhood of the narrow part
of the funnel \citep{Betancourt+Girolami:2019}.  For multimodal
distributions with well-separated modes, the split-\(\widehat{R}\)
adjustment leads to an ESS estimate that is close to the number of
distinct modes that are found. 
In this situation, ESS can be drastically overestimated if computed from a single chain.

A small value of \Rhat\ is not enough to ensure 
that an MCMC sample is useful in practice \citep{vats2018revisiting}.  The effective sample size must also be large enough to get
stable inferences for quantities of interest. \citet{BDA3} proposed an ESS estimate which combines autocovariance-based single-chain variance estimates \citep{Hastings:1970,Geyer:1992} from multiple chains using between- and within-chain information as in \Rhat.
In Section \ref{ESS} we propose an improved algorithm, and
as with \Rhat, we 
recommend computing the ESS on the rank-normalized sample. This does not
directly compute the ESS relevant for computing the mean of the parameter, but 
instead computes a quantity that is well defined even if the chains do not 
have finite mean or variance.  Specifically, it computes the ESS of a sample
from a \emph{rank-normalized} version of the quantity of interest, using the rank transformation followed by the inverse normal transformation. This is still indicative of the effective sample size for computing an average, and if it is low the computed
expectations are unlikely to be good approximations to the actual
target expectations.

To ensure reliable estimates of variances and autocorrelations needed for
\Rhat\ and ESS, we recommend requiring that the rank-normalized ESS is 
greater than $400$, a number we chose based on practical experience and
simulations (see Appendix A) as typically sufficient to get a stable
estimate of the Monte Carlo standard error.

Finally, when reporting quantile estimates or posterior intervals, we 
strongly suggest assessing the convergence of the chains for these quantiles.
In Section \ref{convergence-diagnostics-for-quantiles}, we show that
convergence of Markov chains is not uniform across the parameter space,
that is, convergence might be different in the bulk of the distribution
(e.g., for the mean or median) than in the tails (e.g., for extreme
quantiles). We propose diagnostics and effective sample sizes specifically for 
extreme quantiles. This is different from the standard ESS estimate (which 
we refer to as bulk-ESS), which mainly assesses how well
the centre of the distribution is resolved. Instead, these ``tail-ESS''
measures allow the user to estimate the MCSE for interval estimates.

\section{\Rhat\ and the effective sample size}

When coupled with an ESS estimate,  \Rhat\ is the most common way to
assess the convergence of a set of simulated chains.  There is a link between
these two measures  for a single chain \citep[see, e.g.][]{vats2018revisiting}, but we prefer to 
treat these as two separate questions: ``Did the chains mix well?'' (\sRhat) and 
``Is the effective sample size large enough to get a stable estimate of uncertainty?''
In this section we define the \Rhat\ and ESS statistics that we propose to modify.

%
%

\hypertarget{SplitRhat}{%
\subsection{Split-\Rhat}\label{SplitRhat}}

Here we present split-\(\widehat{R}\),
following \citet{BDA3} but using the notation of
\citet{StanManual.2.18.0}. This formulation represents the current 
standard in convergence diagnostics for iterative simulations. In the
equations below, \(N\) is the number of draws per chain, \(M\) is the
number of chains, \(S=MN\) is the total number of draws from all
chains, \(\theta^{(nm)}\) is $n$th draw of $m$th chain,
\(\theta^{(.m)}\) is the average of draws from $m$th chain, and
\(\theta^{(..)}\) is average of all draws. For each scalar summary of
interest \(\theta,\) we compute \(B\) and \(W,\) the between- and
within-chain variances:
\begin{align}
B &= \frac{N}{M-1}\sum_{m=1}^{M}(\overline{\theta}^{(.m)} - 
\overline{\theta}^{(..)})^2, \quad \mbox{where} \quad 
\overline{\theta}^{(.m)}=\frac{1}{N}\sum_{n=1}^N \theta^{(nm)}, \quad
\overline{\theta}^{(..)} = \frac{1}{M}\sum_{m=1}^M\overline{\theta}^{(.m)} 
\\
W &= \frac{1}{M}\sum_{m=1}^{M}s_m^2, \quad \mbox{where} \quad
s_m^2=\frac{1}{N-1} \sum_{n=1}^N (\theta^{(nm)}-\overline{\theta}^{(.m)})^2.
\end{align}
The between-chain variance, \(B\), also contains the factor \(N\)
because it is based on the variance of the within-chain means,
\(\overline{\theta}^{(.m)},\) each of which is an average of \(N\)
values \(\theta^{(nm)}\). We can estimate \(\var(\theta | y)\),
the marginal posterior variance of the estimand, by a weighted average
of \(W\) and \(B\), namely,
\begin{equation}
\widehat{\var}^+(\theta| y) = \frac{N-1}{N}W + \frac{1}{N}B.
\end{equation}
This quantity \emph{overestimates} the marginal posterior variance
assuming the starting distributions and all intermediate distributions
of the simulations are appropriately
overdispersed compared to the target distribution, but is
\emph{unbiased} under stationarity (that is, if the starting
distribution equals the target distribution), or in the limit
\(N\rightarrow\infty\). To have an overdispersed starting distribution,
independent Markov chains should be initialized with diffuse starting
values for the parameters. 

Meanwhile, for any finite \(N\), the within-chain variance \(W\) should
\emph{underestimate} \(\var(\theta |y)\) because the
individual chains haven't had the time to explore all of the target
distribution and, as a result, will have less variability. In the limit
as \(N\rightarrow\infty\), the expectation of \(W\) also approaches
\(\var(\theta |y)\).

We monitor convergence of the iterative simulations to the target
distribution by estimating the factor by which the scale of the current
distribution for \(\theta\) might be reduced if the simulations were
continued in the limit \(N\rightarrow\infty\). This leads to the estimator
\begin{equation}
\widehat{R} = \sqrt{\frac{\widehat{\var}^+(\theta | y)}{W}},
\end{equation}
which for an ergodic process declines to 1 as \(N\rightarrow\infty\). We call this
split-\(\widehat{R}\) because we are applying it to chains that
have been split in half so that \(M\) is twice the number of simulated
chains. Without splitting, \(\widehat{R}\) would get fooled by
non-stationary chains as in Figure \ref{converge.challenge}b.

In cases, where we can be absolutely certain that a single chain is
sufficient, \(\widehat{R}\) could be computed using only
single chain marginal variance and autocorrelations \citep[see,
e.g.][]{vats2018revisiting}. However we are willing to trade off a
slightly higher variance for increased diagnostic sensitivity (as
described in the introduction) that running multiple chains brings.

\hypertarget{ESS}{%
\subsection{The effective sample size}\label{ESS}}

We estimate effective sample size by combining
information from \Rhat\ and the autocorrelation estimates within the
chains.

\subsubsection*{The effective sample size and Monte Carlo standard error}

Given $S$ independent simulation draws, the accuracy of average of the
simulations \(\bar{\theta}\) as an estimate of the posterior mean
\(\E(\theta | y)\) can be estimated as
\begin{equation}
  \Var(\bar{\theta}) = \frac{\Var(\theta|y)}{S}.
  \label{MCSE}
\end{equation}
This generalizes to posterior expectations of functionals of
parameters \(\E\left(g(\theta) | y\right)\).  The square root of
\eqref{MCSE} is called the Monte Carlo standard error (MCSE).

In general, the simulations of \(\theta\) within each chain
tend to be autocorrelated, and $\Var(\bar{\theta})$ can be larger or
smaller in expectation.
In the early days of using MCMC for Bayesian inference, the focus was
in estimating the single chain estimate variance directly, for
example, based on autocorrelations or batch means
\citep{Hastings:1970,Geyer:1992}. See more different variance
estimation algorithms in reviews by \citet{Cowles+Carlin:1996},
\citet{Mengersen+etal:1999}, and \citet{Robert+Casella:2004}.
Interpreting whether Monte Carlo standard error for a quantity of
interest is small enough requires domain expertise.

Effective sample size (ESS) can be computed by dividing any variance
estimate for an MCMC estimate by the variance estimate assuming
independent draws. As convergence diagnostics in general started to be
more popular
\citep{Gelman+Rubin:1992,Cowles+Carlin:1996,Mengersen+etal:1999,Robert+Casella:2004},
eventually ESS also became popular as description of the efficiency of
the simulation \citep[an early example of reporting ESS for Gibbs
sampler is][]{Sorensen+etal:1995}. The term effective sample size had
already been used before, for example, to describe amount of
information in climatological time series \citep{Laurmann+Gates:1977}
and the efficiency of importance sampling in Bayesian inference
\citep{Kong+Liu+Wong:1994}.

Although ESS is not a replacement for MCSE, it can provide a scale-free
measure of information, which can be
especially useful when diagnosing the sampling efficiency for a large
number of variables. The downside of the term effective sample size is
that it may give a false
impression that the dependent simulation sample would be equivalent to
an independent simulation sample with size ESS, while the equivalence
is only for the estimation efficiency of the posterior mean, and the efficiency of the same
dependent simulation sample for estimating another posterior functional
\(\E\left(g(\theta) | y\right)\) or quantiles can be very
different.
To simplify notation, in this section we consider the effective sample
size for the posterior mean \(\E\left(\theta | y\right)\). This can be 
generalized in a straightforward manner to ESS estimates for \(\E\left(g(\theta) | y\right)\).
Section \ref{convergence-diagnostics-for-quantiles} deals with
estimating the effective sample size of quantiles, which cannot be
presented as expectations.

\subsubsection*{Estimating the effective sample size}

The first proposals of ESS estimates  used information only from
a single chain \citep[see, e.g.][]{Sorensen+etal:1995}. The convergence
diagnostic package \texttt{coda} \citep{coda2006} combines (since
version 0.5.7 in 2001) single chain spectral variance based ESS
estimates simply by summing them, but this approach gives
over-optimistic estimates if spectral variances in different chains
are not equal (e.g. when different step size is used in different
chains) or if chains are not mixing well.  \citet{BDA2} proposed an
ESS estimate,
\begin{equation}
S_{\rm eff,BDA2} = MN\frac{\widehat{\var}^{+}}{B},
\end{equation}
where $\widehat{\var}^+$ is a marginal posterior variance
estimate and $B$ is between-chain variance estimate as given in
Section~\ref{SplitRhat}.  This corresponds to a batch means approach
with each chain being one batch. As there are usually only a small
number of batches (chains), and information from autocorrelations is
not used, this ESS estimate has high variance.
\citet{BDA3} proposed an ESS estimate which appropriately combines
autocorrelation information from multiple chains. \citet{StanManual.2.18.0} made some computational
improvements, and the present article provides a further improved version.



For a single chain of length $N$, the effective sample size of a
chain can defined in terms of the autocorrelations within the chain at
different lags,
\begin{equation}
N_{\rm eff} \ = \
\frac{N}{\sum_{t = -\infty}^{\infty} \rho_t} \ = \
\frac{N}{1 + 2 \sum_{t = 1}^{\infty} \rho_t},
\end{equation}
where \(\rho_t\) is autocorrelation at lag \(t \geq 0\).  An
equivalent approach was used by \citet{Hastings:1970} for
estimating the variance of the mean estimate from a single chain. For a chain
with joint probability function \(p(\theta)\) with mean \(\mu\) and
standard deviation \(\sigma\), \(\rho_t\) is defined to be
\begin{equation}
\rho_t = \frac{1}{\sigma^2} \, \int_{\Theta} (\theta^{(n)} - \mu)
(\theta^{(n+t)} - \mu) \, p(\theta) \, d \theta.
\end{equation}
This is just the correlation between the two chains offset by \(t\)
positions. Because we know \(\theta^{(n)}\) and \(\theta^{(n+t)}\) have
the same marginal distribution at convergence, multiplying the two
difference terms and reducing yields,
\begin{equation}
\rho_t = \frac{1}{\sigma^2} \, \int_{\Theta} \theta^{(n)} \, \theta^{(n+t)}
\, p(\theta) \, d \theta.
\end{equation}

In practice, the probability function in question cannot be tractably
integrated and thus neither autocorrelation nor the effective sample
size can be directly calculated. Instead, these quantities must be
estimated from the sample itself.
Computations of autocorrelations for all lags simultaneously can be
done efficiently via the fast Fourier transform algorithm \citep[FFT;
see][]{Geyer:2011}. In our experiments, FFT-based autocorrelation
estimates have also been computationally more accurate than naive
autocovariance computation. As recommended by \citet{Geyer:1992} we
use the biased estimate with divisor $N$, instead of unbiased
estimate with divisor $N-t$. Also in our experiments, the biased
estimate provided smaller variance in the final ESS estimate.

The autocorrelation estimates \(\hat{\rho}_{t,m}\) at lag \(t\) from
multiple chains \(m \in (1,\ldots,M)\) are combined with the
within-chain variance estimate \(W= \frac{1}{M}\sum_{m=1}^{M}s_m^2\)
and the multi-chain variance estimate
\(\widehat{\var}^{+} = W(N-1)/N+B/N\) to compute the combined
autocorrelation at lag \(t\) as,
\begin{equation}
\hat{\rho}_t
= 1 - \frac{\displaystyle W - \textstyle \frac{1}{M} \sum_{m=1}^M
s_m^2 \hat{\rho}_{t,m}}{\widehat{\var}^{+}}. \label{rhohat}
\end{equation}
If $\hat{\rho}_{t,m}=0$ for all $m$, $\hat{\rho}_t=1-\widehat{R}^{-2}$.
If in addition chains are mixing well so that
$\widehat{R}\approx 1$, then $\hat{\rho}_t \approx 0$. If
$\hat{\rho}_{t,m} \neq 0$ and $\widehat{R} \approx 1$, then
$\hat{\rho}_t \approx \frac{1}{M} \sum_{m=1}^M \hat{\rho}_{t,m}$. If
$\widehat{R} \gg 1$, then $\hat{\rho}_t \approx 1-\widehat{R}^{-2}$.
If chains are mixing well, this expression is equivalent to averaging autocorrelations, and if
chains are not mixing well, simulations in each chain are implicitly assumed to be
more correlated with each other.
In our experiments, multi-chain $\rho_t$ given by \eqref{rhohat} and FFT-based $\hat{\rho}_{t,m}$ had smaller variance than the related
multi-chain $\rho_t$ proposed by \citet{BDA3}.

As noise in the correlation estimates \(\hat{\rho}_t\) increases as
\(t\) increases, the large-lag terms need to be down weighted
\citep[lag window approach, see,
e.g.][]{Geyer:1992,Flegal+Jones:2010} or the sum of
\(\hat{\rho}_t\) can be truncated with some truncation lag $T$ to get
\begin{equation}
S_{\rm eff} \ = \ \frac{NM}{1 + 2 \sum_{t = 1}^{T} \rho_t}.
\end{equation}
We use a truncation rule proposed by \citet{Geyer:1992}, which takes
into account certain properties of the autocorrelations for Markov
chains.  Even when the simulations are constructed using an MCMC
algorithm, the time series of simulations for a scalar parameter or
summary will not in general have the Markov property; nonetheless we
have found these Markov-derived heuristics to work well in practice.
In our experiments, Geyer's truncation had superior stability
compared to flat-top \citep[e.g.][]{Doss+etal:2014:MCMC-quantiles} and
slug-sail \citep{vats2018revisiting} lag window approaches.

For Markov chains typically used in MCMC, negative autocorrelations
can happen only on odd lags and by summing over pairs starting from
lag \(t=0\), the paired autocorrelation is guaranteed to be positive,
monotone and convex modulo estimator noise \citep{Geyer:1992,
  Geyer:2011}. The effective sample size of combined chains is then
defined as
\begin{equation}
S_{\rm eff} = \frac{N \, M}{\hat{\tau}},
\end{equation}
where
\begin{equation}
\hat{\tau} = 1 + 2 \sum_{t=1}^{2k+1} \hat{\rho}_t = 
-1 + 2 \sum_{t'=0}^{k} \hat{P}_{t'},
\end{equation}
and \(\hat{P}_{t'}=\hat{\rho}_{2t'}+\hat{\rho}_{2t'+1}\). The initial
positive sequence estimator is obtained by choosing the largest \(k\)
such that \(\hat{P}_{t'}>0\) for all \(t' = 1,\ldots,k\). The initial
monotone sequence estimator is obtained by further reducing
\(\hat{P}_{t'}\) to the minimum of the preceding values so that the
estimated sequence becomes monotone.

In case of antithetic Markov chains, which have negative
autocorrelations on odd lags, the effective sample size
\(S_{\rm eff}\) can also be larger than \(S\) . For example, the
dynamic Hamiltonian Monte Carlo algorithms used in Stan
\citep{Hoffman+Gelman:2014, betancourt2017conceptual, StanManual.2.18.0} is likely to
produce \(S_{\rm eff}>S\) for parameters with a close to Gaussian
posterior (in the unconstrained space) and low dependence on the other
parameters.
The benefit of this kind of super-efficiency is often
limited as it is unlikely to simultaneously have super-efficiency for
mean and variance (or tail quantiles) as demonstrated in our experiments.

In extreme antithetic cases, magnitude of single lag autocorrelations
can stay large for a large lag $t$, even if the paired
autocorrelations are close to zero. To improve the stability and
reduce the variance of the ESS estimate, we determine the truncation
lag as usual, but compute the average of truncated sum ending to usual
odd lag and truncated sum ending to the next even lag.
Sometimes these estimates are used for very short antithetic chains,
and just by chance there can be strange estimates, and as highly
antithetic chains are unlikely, in our software implementation we have
restricted the ESS estimate to an upper bound of $S\log_{10}(S)$.

The effective sample size \(S_{\rm eff}\) described here is different
from similar formulas in the literature in that we use multiple chains
and between-chain variance in the computation, which typically gives us
more conservative claims (lower values of \(S_{\rm eff}\)) compared to
single chain estimates, especially when mixing of the chains is poor. If
the chains are not mixing at all (e.g., if the posterior is multimodal and
the chains are stuck in different modes), then our \(S_{\rm eff}\) is
close to the number of distinct modes that are found. Thus, our ESS estimate can
be also to diagnose multimodality.


The values of \Rhat\ and ESS require reliable estimates of variances and autocorrelations
(in addition to the existence of these quantities; see our Cauchy examples in Section 
\ref{cauchy-a-distribution-with-infinite-mean-and-variance}), which can
only occur if the chains have enough independent replicates. In particular, we only recommend 
relying on the \Rhat\ estimate to make decisions about the quality of the chain if each of the 
split chains has an average ESS estimate of at least $50$. In our minimum recommended setup of four
parallel chains, the total ESS should be at least $400$ before we expect  \Rhat\ to be useful.

\hypertarget{improving-convergence-diagnostics}{%
\section{Improving convergence
diagnostics}\label{improving-convergence-diagnostics}}

\hypertarget{rank-normalization}{%
\subsection{Rank normalization helps  \Rhat\ when there are heavy tails}\label{rank-normalization}}

As split-\(\widehat{R}\) and \(S_{\rm eff}\) are well defined
only if the marginal posteriors have finite mean and variance, we
propose to use rank normalized parameter values instead of the actual
parameter values for the purpose of diagnosing convergence.

The use of ranks to avoid the assumption of normality goes back to
\citet{Friedman:1937}. \citet{Chernoff+Savage:1958} show rank based
approaches have good asymptotic efficiency. Instead of using rank
values directly and modifying tests for them,
\citet{Fisher+Yates:1938} propose to use expected normal scores
(ordered statistics) and use the normal models. \citet{Blom:1958}
shows that accurate approximation of the expected normal scores can
be computed efficiently from ranks using an inverse normal transformation.

Rank normalized split-\(\widehat{R}\) and \(S_{\rm eff}\) are
computed using the equations in Section \ref{SplitRhat} and \ref{ESS}, but
replacing the original parameter values \(\theta^{(nm)}\) with their
corresponding rank normalized values (normal scores) denoted as \(z^{(nm)}\). Rank
normalization proceeds as follows.  First, replace each value
\(\theta^{(nm)}\) by its rank \(r^{(nm)}\) within the pooled draws
from all chains.  Average rank for ties are used to conserve the
number of unique values of discrete quantities.  Second, transform
ranks to normal scores using the inverse normal transformation and
a fractional offset \citep{Blom:1958}:
\begin{equation}
z^{(nm)} = \Phi^{-1}\left(\frac{r^{(nm)}-3/8}{S+1/4}\right).
\end{equation}
Using normalized ranks (normal scores)
\(z^{(nm)}\) instead of ranks \(r^{(nm)}\) themselves has the
benefits that (1) for continuous variables the normality assumptions in computation of \(\widehat{R}\) and \(S_{\rm eff}\) are fulfilled (via the transformation), (2) the values of \(\widehat{R}\) and \(S_{\rm eff}\) are practically the
same as before for nearly normally distributed variables (the interpretation doesn't change for the cases where the original \(\widehat{R}\) worked well), and (3) rank-normalized 
\Rhat\ and \(S_{\rm eff}\) are invariant to monotone transformations (e.g. we get the same diagnostic values when examining a variable or logarithm of a variable).
The effects of rank normalization are further explored in the online appendix.

We will use the term \emph{bulk effective sample size} (bulk-ESS or
bulk-\(S_{\rm eff}\)) to refer to the effective sample size based on the
rank normalized draws. Bulk-ESS is useful for diagnosing problems due to
trends or different locations of the chains (see Appendix~A). Further, it is
well defined even for distributions with infinite mean or variance, a
case where previous ESS estimates fail. However, due to the rank
normalization, bulk-ESS is no longer directly applicable to estimate the
Monte Carlo standard error of the posterior mean. We will come back to
the issue of computing Monte Carlo standard errors for relevant
quantities in Section \ref{mcse}.

\hypertarget{diagnostics-for-folded-draws}{%
\subsection{Folding reveals problems with variance and tail exploration}\label{diagnostics-for-folded-draws}}

Both original and rank normalized split-\(\widehat{R}\) can be
fooled if the chains have the same location but different scales. This
can happen if one or more chains is stuck near the middle of the distribution. 
To alleviate this problem, we propose a
rank normalized split-\(\widehat{R}\) statistic not only for the
original draws \(\theta^{(nm)}\), but also for the corresponding {\em folded}
draws \(\zeta^{(mn)}\), absolute deviations from the median,
\begin{equation}
\label{zeta}
\zeta^{(mn)} = \left|\theta^{(nm)}-{\rm median}(\theta)\right|.
\end{equation}
We call the rank normalized split-\(\widehat{R}\) measure computed on the
 \(\zeta^{(mn)}\) values  \emph{folded-split}-\(\widehat{R}\).
  This measures convergence in the
tails rather than in the bulk of the distribution. To obtain a single
conservative \(\widehat{R}\) estimate, we propose to report the maximum
of rank normalized split-\(\widehat{R}\) and rank normalized
folded-split-\(\widehat{R}\) for each parameter.

Figure \ref{converge.challenge} demonstrates how our new version  of \Rhat\  catches some examples of lack of convergence that were not detected by earlier versions of the potential scale reduction factor. We do not intend with this example to claim that our new \Rhat\ is perfect---of course, it can be defeated too.  Rather, we use these simple scenarios to develop intuition about problems with  traditional  \sRhat\  and possible directions for improvement.

\hypertarget{convergence-diagnostics-for-quantiles}{%
\subsection{Localizing convergence diagnostics: assessing the quality of quantiles, the median absolute deviation, and small-interval probabilities }\label{convergence-diagnostics-for-quantiles}}

The new \(\widehat{R}\) and bulk-ESS introduced above are useful as overall efficiency
measures. Next we introduce
convergence diagnostics for quantiles and related quantities, which are more focused measures and help to diagnose reliability of  reported
posterior intervals. Estimating
the efficiency of quantile estimates has a high practical
relevance in particular as we observe the efficiency for tail quantiles
to often be lower than for the mean or median.  This especially has implications
if people are making decisions based on whether or not a specific quantile
is below or above a fixed value (for example, if a posterior interval contains zero).

The \(\alpha\)-quantile
is defined as the parameter value \(\theta_\alpha\) for which
\(\Pr(\theta \leq \theta_\alpha) = \alpha\). An estimate
\(\hat{\theta}_\alpha\) of \(\theta_\alpha\) can be obtained by
finding the \(\alpha\)-quantile of the empirical cumulative distribution function (ECDF) of the
posterior draws \(\theta^{(s)}\).

The cumulative probabilities \(\Pr(\theta \leq \theta_\alpha)\) can be
written as expectation which can be estimated with sample mean
\begin{equation}
\Pr(\theta \leq \theta_\alpha) = \E(\I(\theta \leq \theta_\alpha)) \approx \bar{\I}_\alpha = \frac{1}{S}\sum_{s=1}^S
\I(\theta^{(s)} \leq\theta_\alpha),
\end{equation}
where \(\I(\cdot)\) is the indicator function. The indicator function
transforms simulation draws to 0's and 1's, and thus the subsequent
computations are bijectively invariant.
Efficiency estimates of the ECDF at any \(\theta_\alpha\) can now be
obtained by applying rank-normalizing and subsequent computations
directly on the indicator function's results.
More details on the variance of the cumulative
distribution function can be found in the online appendix.
\citet{Raftery+Lewis:1992a} proposed to focus on accuracy of
cumulative or interval probabilities and also proposed a specific effective
sample size estimate for these probability estimates.

Although the quantiles cannot be written directly as an expectation,
the quantile estimate is strongly consistent and
\citet{Doss+etal:2014:MCMC-quantiles} provide conditions for a
quantile central limit theorem.
Assuming that the CDF is a continuous function \(F\) which is smooth
near an \(\alpha\)-quantile of interest, we could compute 
\begin{equation}
  \Var(\hat{\theta}_\alpha) = \Var(F^{-1}(\bar{\I}_\alpha)) = \Var(\bar{\I}_\alpha)/f(\theta_\alpha).
\end{equation}
Even if we do not usually know \(F\), this shows that the variance of
\(\theta_\alpha\) is just the variance of \(\bar{\I}_\alpha\) scaled by
the unknown density \(f(\theta_\alpha)\), and thus the
effective sample size for the quantile estimate
\(\hat{\theta}_\alpha\) is the same as for the corresponding
cumulative probability.

To get a better sense of the sampling efficiency in the
distributions' tails, we propose to compute the minimum of the effective
sample sizes of the 5\% and 95\% quantiles, which we will call
\emph{tail effective sample size} (tail-ESS or tail-\(S_{\rm eff}\)).
Tail-ESS can help diagnosing problems due to different scales of the
chains (see Appendix~A).

Since the marginal posterior distributions might not have finite mean
and variance, for example, the popular \texttt{rstanarm} package
\citep{RStanARM.2.17} reports median and median absolute deviation (MAD)
instead of mean and standard error. Median and MAD are well defined
even when the marginal distribution does not have finite mean and
variance. Since the median is same as the 50\% quantile, we can get an
efficiency estimate for it as for any other quantile.

Further, we can also compute an efficiency estimate for the median
absolute deviation by computing the efficiency estimate of an
indicator function based on the folded parameter values \(\zeta\) (see
(\ref{zeta})):
\begin{equation}
\Pr(\zeta \leq \zeta_{0.5}) \approx \bar{\I}_{\zeta,0.5} = \frac{1}{S}\sum_{s=1}^S
\I(\zeta^{(s)} \leq \zeta_{0.5}),
\end{equation}
where \(\zeta_{0.5}\) is the median of the folded values. The efficiency estimate for the MAD is obtained by applying the same
approach as for the median (and other quantiles) but with the folded
parameters values.

We can get more local efficiency estimates by considering small
probability intervals. We propose to compute the efficiency estimates
for
\begin{equation}
\bar{\I}_{\alpha,\delta} = \Pr(\hat{Q}_\alpha < \theta \leq \hat{Q}_{\alpha+\delta}),
\end{equation}
where \(\hat{Q}_\alpha\) is an empirical \(\alpha\)-quantile,
\(\delta=1/k\) is the length of the interval for some positive integer
\(k\), and \(\alpha \in (0,\delta,\ldots,1-\delta)\) changes in steps of
\(\delta\). Each interval has \(S/k\) draws, and the efficiency measures
the autocorrelation of an indicator function which is \(1\) when the
values are inside the specific interval and \(0\) otherwise. This
gives us a local efficiency measure which is more localized than
efficiency measure for quantiles and can be used to build intuition
about what types of posterior functionals can be computed as
illustrated in the examples. While the expectation of a function that
only depends on intermediate values can be usually estimated with relative
ease, expectations of tail probabilities or other posterior
functionals that depend critically on the tail of the
distribution will be usually more difficult to estimate. In addition, small
probability intervals can be used in practical equivalence testing
\citep[see, e.g.,][]{Wellek:2010:testing}.

A natural multivariate extension of small intervals would be to consider
small probability volumes using a box or sphere with dimensions
determined, for example, by marginal quantiles. The visualization of
the multivariate results would be easiest in 2 or 3 dimensions. In
higher dimensions, for example, $k$-means clustering could be used to
determine hyper-spheres. Even if it gets more difficult to visualize
where the problematic region in the high dimensional space
is, the diagnosing that sampling efficiency is low in some parts of the
posterior can be useful.

\hypertarget{mcse}{%
\subsection{Monte Carlo error estimates for quantiles}\label{mcse}}

To obtain the MCSE for
\(\hat{\theta}_\alpha\), \citet{Doss+etal:2014:MCMC-quantiles} use a
Gaussian kernel density estimate of \(f(\theta_\alpha)\) and
batch means and subsampling bootstrap method for estimating
\(\Var(\bar{\I}_\alpha)\), and \citet{Liu+etal:2016:MCMC-quantiles} use a
flat top kernel density estimate for \(f(\theta_\alpha)\) and a
spectral variance approach for \(\Var(\bar{\I}_\alpha)\).

We propose an alternative approach which avoids the need
to estimate \(f(\theta_\alpha)\). Here is how we estimate, for example,
a central 90\% Monte Carlo error interval for \(\hat{\theta}_\alpha\)
(any quantiles or intervals can be computed using the same algorithm):
\begin{enumerate}
\def\labelenumi{\arabic{enumi}.}
\item Compute the effective sample size \(S_{\rm eff}\) for estimating the expectation \(\E(\I(\theta \leq \hat{\theta}_\alpha))\).
\item Compute \(a\) and \(b\) as \(5\%\) and \(95\%\) quantiles (for other than \(90\%\) interval use corresponding quantiles) of
  \begin{equation}
  \Beta\left(S_{\rm eff} \alpha+1,\, S_{\rm eff} (1-\alpha) + 1\right).
\end{equation}
  Using $S_{\rm eff}$ here takes into account the efficiency of the
  posterior draws. The variance of this beta distribution matches the
  variance of normal approximation, but using quantiles guarantees
  that $0<a<1$ and $0<b<1$.  Asymptotically as
  \(S_{\rm eff} \rightarrow \infty \), this beta distribution
  converges towards a normal distribution. Instead of drawing random
  sample from the beta distribution, we get sufficient accuracy for
  MCSE using just two deterministically chosen quantiles.
\item Propagate $a$ and $b$ through the nonlinear inverse transforms
  $A=(F^{-1}(a))$ and $B=(F^{-1}(b))$. Then $A$ and $B$ are
  corresponding quantiles in the transformed scale.
  As we don't know $F$ for the quantity of interest, we use a simple numerical approximation:
  \begin{align*}
  \widehat{A} & = \theta^{(s')} \,\,\text{ where }\,  s' \leq Sa < s'+1 \\
  \widehat{B} & = \theta^{(s'')}  \,\text{ where }\,  s''-1 < Sb \leq s'',
  \end{align*}
  where $\theta^{(s)}$ have been sorted in ascending order.
  $\widehat{A}$ and $\widehat{B}$ are then estimated \(5\%\) and
  \(95\%\) quantiles (or other quantiles corresponding to which
  quantiles $a$ and $b$ were chosen to be) of the Monte Carlo error
  interval for \(\hat{\theta}_\alpha\).
\end{enumerate}

The Monte Carlo standard error for \(\hat{\theta}_\alpha\) can be
approximated, for example, by computing $(\widehat{B}-\widehat{A})/2$,
where $\widehat{A}$ and $\widehat{B}$ are estimated \(16\%\) and
\(84\%\) Monte Carlo error quantiles computed with the above
algorithm.
Use of deterministically chosen 16\% and 84\% quantiles $a$ and $b$,
propagating them through the nonlinear transformation and estimating
the standard error from the transformed quantiles, corresponds to
unscented transformation which is known to estimate the variance of
the transformed quantity correct to the second order
\citep{Julier+Uhlman:1997:unscented}.

The above algorithm is useful as a default, as it is more robust than
density estimation based approaches for
non-smooth densities, which is common case, for example, when
variables are constrained in a (semi-open) range.  $\widehat{A}$ and $\widehat{B}$ are likely
to have high variance in case of extreme tail quantiles and thick-tailed distributions,
as there are not many $\theta^{(s)}$ in extreme
tails. The approaches using a density estimate for \(f(\theta_\alpha)\)
can provide better accuracy when the assumptions of the density estimate
are fulfilled, but they can have a high bias if the density is not
smooth or the shape of the kernel doesn't match well the tail
properties of the distribution. To improve accuracy of extreme tail
quantile estimates, common extreme value models could be used to model
the tail of the distribution.

\hypertarget{diagnostic-visualizations}{%
\subsection{Diagnostic visualizations}\label{diagnostic-visualizations}}

In order to develop intuitions around the convergence of iterative algorithms, we
propose several new diagnostic visualizations in addition to the numerical
convergence diagnostics discussed above. We illustrate with several examples in Section
\ref{examples}.

\begin{figure}[tp]
  \centering
  \begin{minipage}{0.48\textwidth}
  \includegraphics[width=0.98\textwidth]{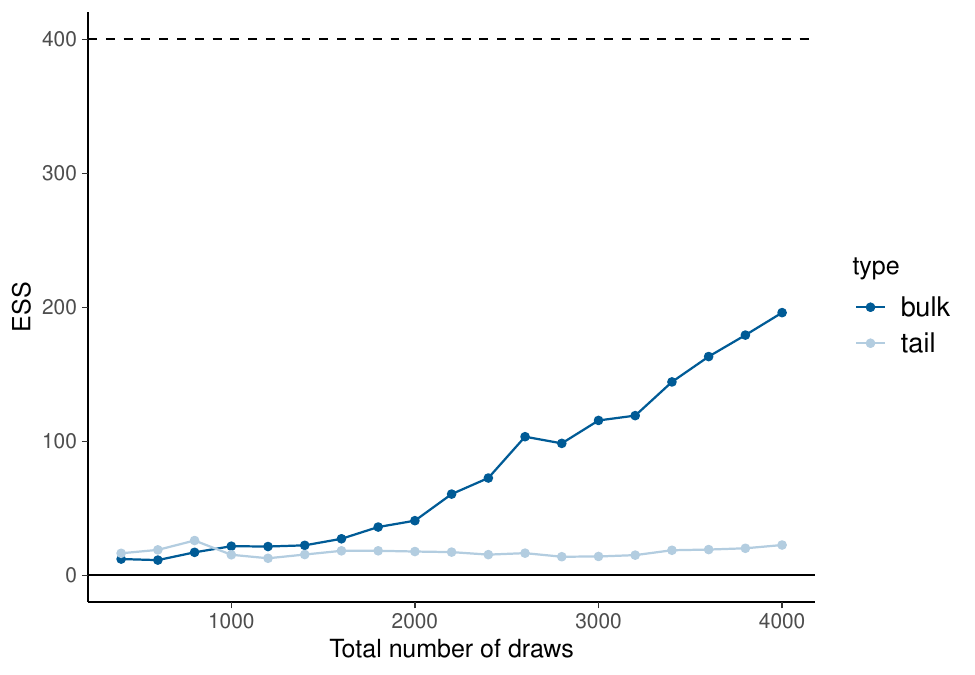}
  \caption{Estimated effective sample sizes with increasing number of iterations
  for the Cauchy model with nominal parameterization. Results are displayed 
  for the element of $x$ with the smallest tail-ESS. The dashed line shows the
    recommended threshold of $400$.}
  \label{fig:change-ess-fit-nom-1}
\end{minipage}
\hfill
\begin{minipage}{0.48\textwidth}
  \includegraphics[width=0.98\textwidth]{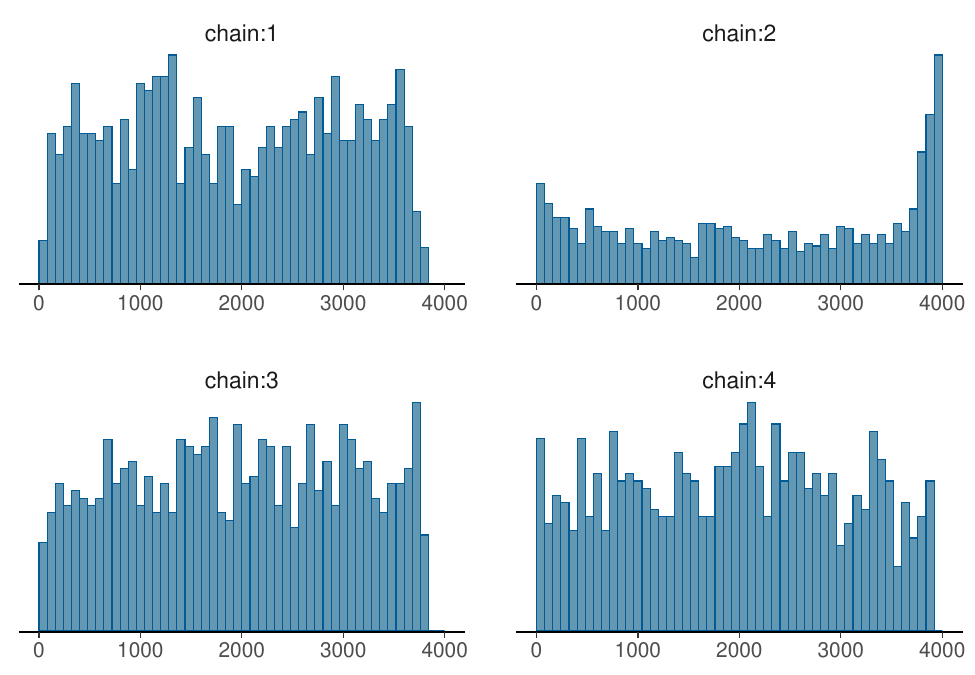}
  \caption{Rank plots of posterior draws from four chains for the Cauchy model 
  with nominal parameterization. Results are displayed for the element of 
  $x$ with the smallest tail-ESS.}
  \label{fig:hist-fit-nom-1}
\end{minipage}
\end{figure}

\hypertarget{rank-plots}{%
\paragraph{Rank plots.}\label{rank-plots}}
Extending the idea of using ranks instead of the original parameter
values, we propose using rank plots for each chain instead
of trace plots. Rank plots, such as Figure \ref{fig:hist-fit-nom-1}, are histograms of the
ranked posterior draws (ranked over all chains) plotted separately for
each chain. If all of the chains are targeting the same posterior, we expect the 
ranks in each chain to be uniform, whereas if one chain has a different location
or scale parameter, this will be reflected in the deviation from uniformity. 
 If rank plots of all chains look similar, this indicates
good mixing of the chains. As compared to trace plots, rank plots don't
tend to squeeze to a fuzzy mess when used with long chains.

\hypertarget{quantile-and-small-interval-plots}{%
\paragraph{Quantile and small-interval
plots.}\label{quantile-and-small-interval-plots}}
The efficiency of quantiles or small-interval probabilities may vary
drastically across different quantiles and small-interval positions,
respectively. We thus propose to use diagnostic plots that display
efficiency of quantiles or small-interval probabilities across their
whole range to better diagnose areas of the distributions that the
iterative algorithm fails to explore efficiently.

\hypertarget{efficiency-change-plots}{%
\paragraph{Efficiency per iteration plots.}\label{efficiency-change-plots}}
For a well-explored distribution, we expect the ESS measures to grow
linearly with the total number of draws \(S\), or, equivalently, that
the relative efficiency (ESS divided \(S\)) is approximately constant
for different values of \(S\). For small number of draws, both bulk and
tail-ESS may be unreliable and cannot necessarily reveal convergence
problems. As a result, some issues may only be
detectable as \(S\) increases, if ESS grows sublinearly
or even decreases with increasing \(S\). Equivalently, in
such a case, we would expect to see a relatively sharp drop in the
relative efficiency measures. We therefore propose to plot the change of
both bulk and tail ESS with increasing \(S\). This can be done based on
a single model without a need to refit, as we can just extract initial
sequences of certain length from the original chains. However, some
convergence problems only occur at relatively high
\(S\) and may thus not be detectable if the total number of draws is too
small.

\hypertarget{examples}{%
\section{Examples}\label{examples}}

We now demonstrate our approach and recommended workflow on several small examples.
Unless mentioned otherwise, we use dynamic Hamiltonian Monte Carlo with
multinomial sampling \citep{betancourt2017conceptual} as implemented
in Stan \citep{StanManual.2.18.0}. We run 4 chains, each with 1000
warmup iterations, which do not form a Markov chain and are discarded, and
1000 post-warmup iterations, which are saved and used for inference.

\hypertarget{cauchy-a-distribution-with-infinite-mean-and-variance}{%
\subsection{Cauchy: A distribution with infinite mean and
variance}\label{cauchy-a-distribution-with-infinite-mean-and-variance}}

Traditional \(\widehat{R}\) is based on calculating
within and between chain variances. If the marginal distribution of a quantity of interest
is such that the variance is infinite,
this approach is not well justified, as we demonstrate here with a Cauchy-distributed example.

\hypertarget{nominal-parameterization-of-the-cauchy-distribution}{%
\subsubsection*{Nominal parameterization of the
Cauchy distribution}\label{nominal-parameterization-of-the-cauchy-distribution}}

We start by simulating  from  independent standard Cauchy distributions for each 
element of a 50-dimensional vector $x$:

\begin{equation}
x_j \sim \text{Cauchy}(0, 1)  \quad \text{for} \quad j=1,\dots,50.
\end{equation}

\begin{figure}[tp]
  \centering
  \begin{minipage}{0.48\textwidth}
  \includegraphics[width=0.98\textwidth]{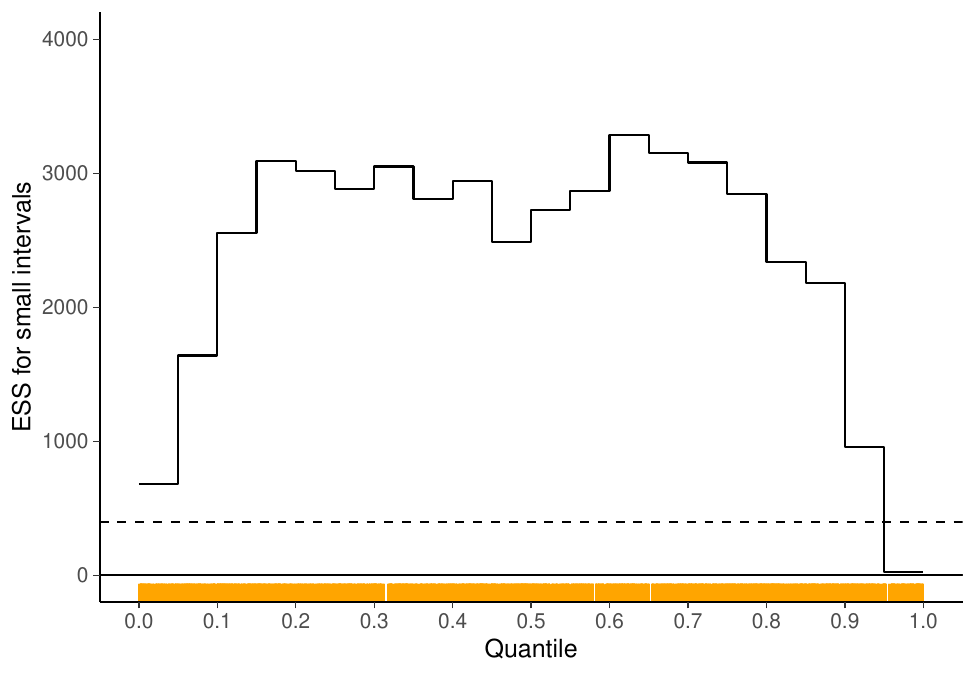}
  \caption{Local efficiency of small-interval probability
    estimates for the Cauchy model with nominal parameterization. 
    Results are displayed for the element of $x$ with the smallest tail-ESS. The dashed line shows the
    recommended threshold of $400$.
    Orange ticks show the position of iterations that exceeded the maximum 
    tree depth in the dynamic HMC algorithm.}
\label{fig:local-ess-fit-nom-1}
\end{minipage}
\hfill
  \begin{minipage}{0.48\textwidth}
  \includegraphics[width=0.98\textwidth]{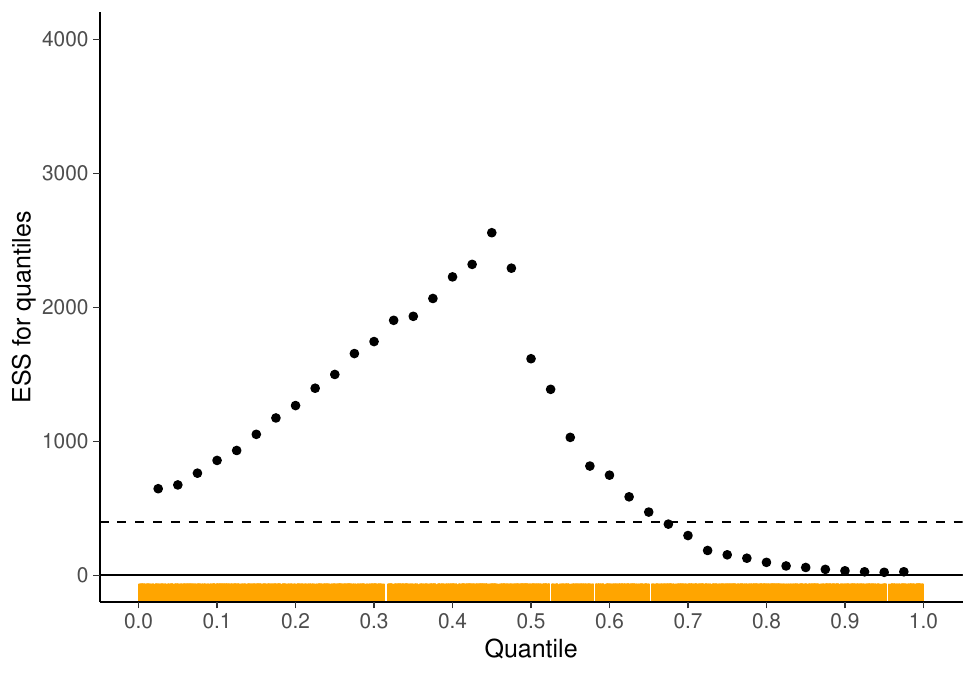}
  \caption{Efficiency of quantile estimates for the Cauchy model with nominal 
  parameterization. Results are displayed for the element of $x$ with the 
  smallest tail-ESS. The dashed line shows the
    recommended threshold of $400$. Orange ticks show the position of iterations that 
  exceeded the maximum tree depth in the dynamic HMC algorithm.\\~}
  \label{fig:quantile-ess-fit-nom-1}
\end{minipage}
\end{figure}

We monitor the convergence for each of the $x_j$ separately. As the distribution of $x$
has thick tails, we may expect any generic MCMC algorithm to have mixing problems.
Several values of \Rhat\ greater than 1.01 and some effective sample sizes less 
than 400 also indicate convergence problems (in addition a HMC-specific diagnostic,
``iterations exceed maximum tree depth'' \citep{StanManual.2.18.0} also
indicated slow mixing of the chains).
The online appendix contains more results 
with longer chains and other \(\widehat{R}\) diagnostics.
We can further analyze potential problems using local efficiency and
rank plots. We specifically investigate \(x_{36}\), which, in this
specific run, had the smallest tail-ESS of 34. 
Figure~\ref{fig:local-ess-fit-nom-1} shows the local efficiency of small 
interval probability estimates (see Section \ref{convergence-diagnostics-for-quantiles}).
The efficiency of sampling is low in the tails, which is clearly
caused by slow mixing in long tails of the Cauchy distribution.  
Figure~\ref{fig:quantile-ess-fit-nom-1} shows the efficiency
of quantile estimates (see Section \ref{convergence-diagnostics-for-quantiles}), 
which also is low in the tails. 

We may also investigate how the estimated effective sample sizes
change when we use more and more draws; \citet{Brooks+Gelman:1998}
proposed to use similar graph for \(\widehat{R}\). If the effective
sample size is highly unstable, does not increase proportionally with
more draws, or even decreases, this indicates that simply running
longer chains will likely not solve the convergence issues. In
Figure~\ref{fig:change-ess-fit-nom-1}, we see how unstable both
bulk-ESS and tail-ESS are for this example.
Rank plots in Figure~\ref{fig:hist-fit-nom-1} clearly show the
mixing problem between chains. In case of good mixing all rank plots
should be close to uniform. More experiments can be found in Appendix~B 
and in the online appendix.

\hypertarget{alternative-parameterization-of-the-cauchy-distribution}{%
\subsubsection*{Alternative parameterization of the
Cauchy distribution}\label{alternative-parameterization-of-the-cauchy-distribution}}

Next, we examine an alternative parameterization  of the
Cauchy as a scale mixture of Gaussians:
\begin{align}
  a_j \sim  \N(0,1), \qquad
  b_j \sim  \Gam (0.5, 0.5), \qquad
  x_j =  a_j/\sqrt{b_j}.
\end{align}
The model has two parameters which have thin-tailed distributions so that we may
assume good mixing of Markov chains. Cauchy-distributed \(x\) can be
computed deterministically from \(a\) and \(b\). In addition to improved sampling 
performance, the example illustrates that focusing on diagnostics matters.
We define two 50-dimensional parameter vectors $a$ and $b$ from which
the 50-dimensional quantity $x$ is computed.

\begin{figure}[tp]
  \centering
  \begin{minipage}{0.48\textwidth}
  \includegraphics[width=0.98\textwidth]{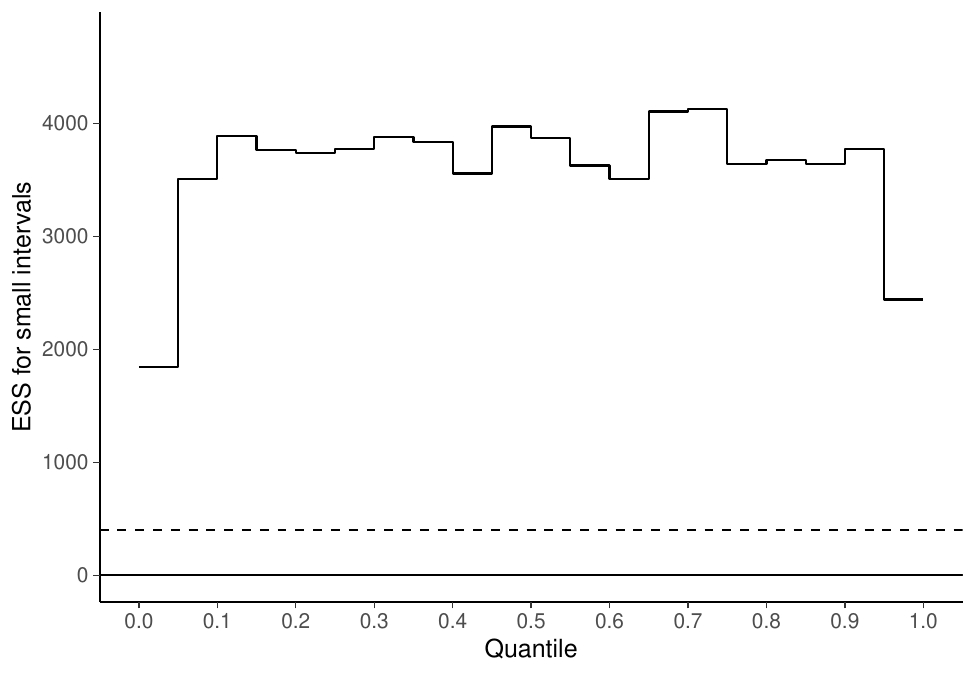}
  \caption{Local efficiency of small-interval probability estimates for the 
  Cauchy model with alternative parameterization. Results are displayed for
  the element of $x$ with the smallest tail-ESS. The dashed line shows the
    recommended threshold of $400$.}
\label{fig:local-ess-fit-alt1-1}
\end{minipage}
\hfill
  \begin{minipage}{0.48\textwidth}
  \includegraphics[width=0.98\textwidth]{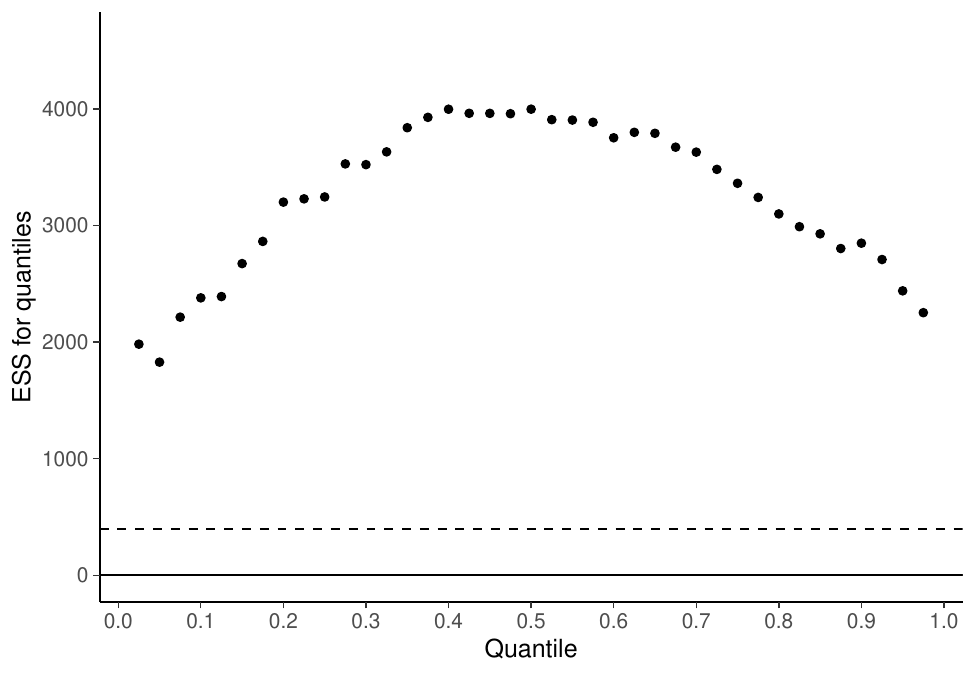}
  \caption{Efficiency of quantile estimates for the Cauchy model with 
  alternative parameterization. Results are displayed for the element of $x$
  with the smallest tail-ESS. The dashed line shows the
    recommended threshold  of $400$.}
  \label{fig:quantile-ess-alt1-1}
\end{minipage}
\end{figure}

\begin{figure}[tp]
  \centering
  \includegraphics[width=0.47\textwidth]{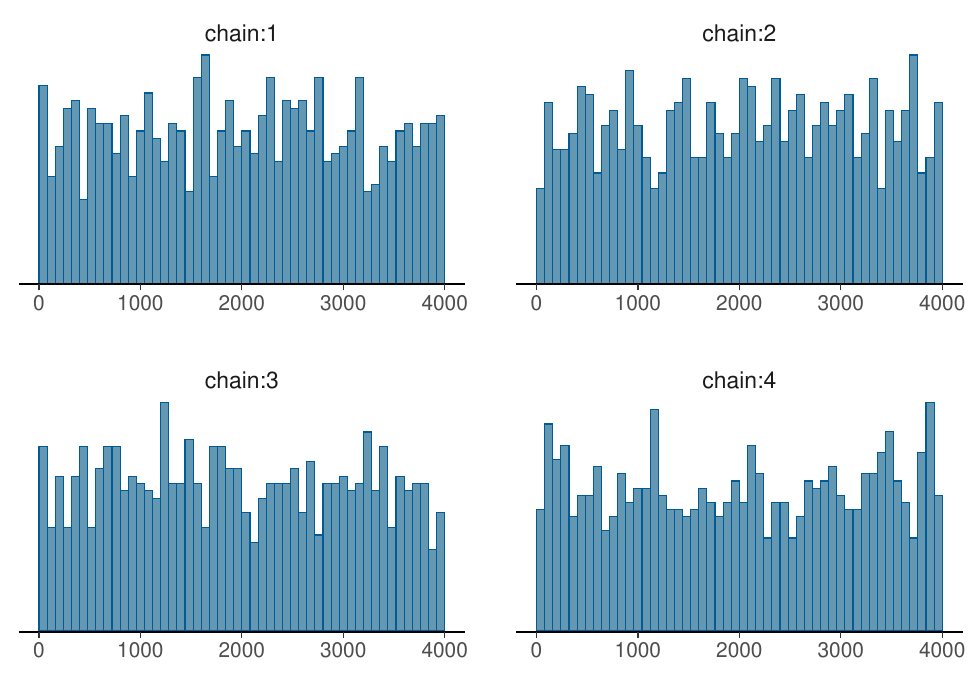}
  \caption{Rank plots of posterior draws from four chains for the Cauchy 
  model with alternative parameterization. Results are displayed for the
  element of $x$ with the smallest tail-ESS.}
  \label{fig:hist-fit-alt1-1}
\end{figure}

For all parameters, \Rhat\ is less than \(1.01\) and 
ESS exceeds 400, indicating that
sampling worked much better with this alternative parameterization.
The online appendix contains more results using other parameterizations 
of the Cauchy distribution. The vectors \(a\) and \(b\) used
to form the Cauchy-distributed \(x\) have stable quantile, mean and
variance values. The quantiles of each \(x_j\) are stable too, but the mean 
and variance estimates are widely varying.
We can further analyze potential problems using local efficiency
estimates and rank plots. For this example, we take a detailed look at 
\(x_{40}\), which had the smallest bulk-ESS of 2848.
Figures~\ref{fig:local-ess-fit-alt1-1} and
\ref{fig:quantile-ess-alt1-1} show good sampling efficiency for the
small-interval probability and quantile estimates.
The rank plots in Figure~\ref{fig:hist-fit-alt1-1} also look close to
uniform across chains, which is consistent with good mixing.
The appearances of the plots in Figures \ref{fig:local-ess-fit-alt1-1},
\ref{fig:quantile-ess-alt1-1}, and \ref{fig:hist-fit-alt1-1} are what
we would expect for well mixing chains in general.

\begin{figure}[htb]
\centering
  \begin{tabular}{@{}cccc@{}}
    \includegraphics[width=.49\textwidth]{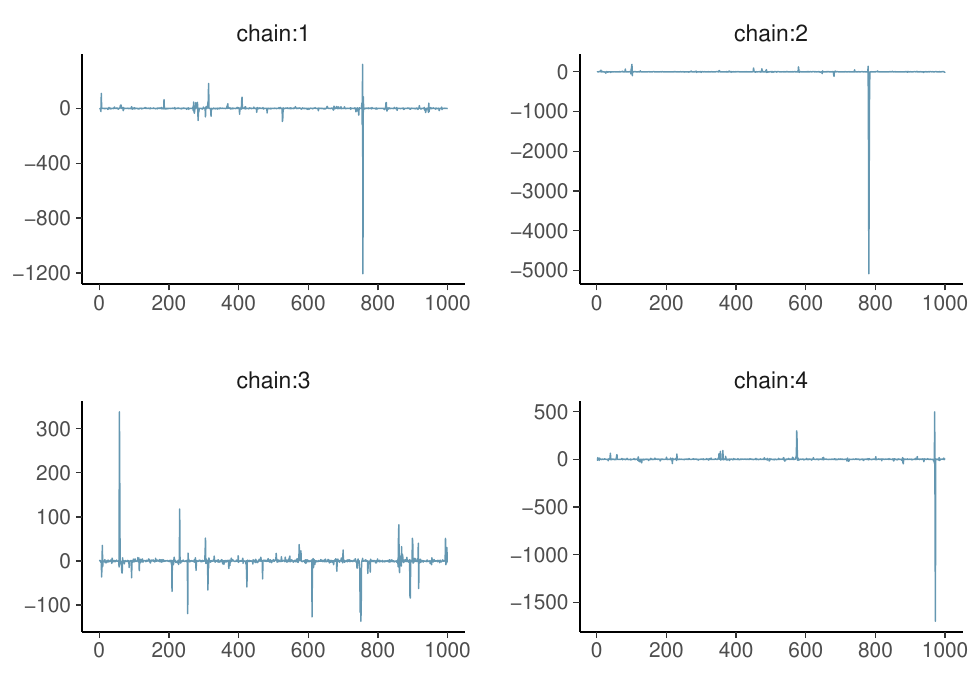} &
    \includegraphics[width=.49\textwidth]{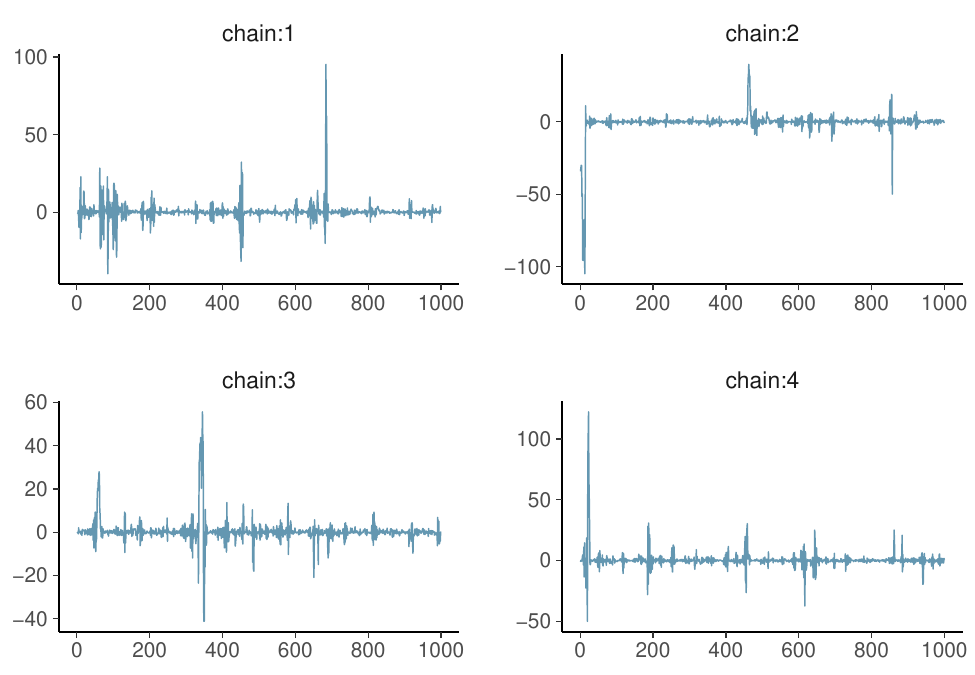}
  \end{tabular}
  \caption{Trace plots of posterior draws from four chains for the Cauchy 
  model with nominal and alternative parameterization. We do not tell which plot belongs to which model and let the reader decide themselves how easy it is to see differences in convergence from those trace plots. Results are displayed for the
  element of $x$ with the smallest tail-ESS in the respective model.}
  \label{fig:trace-cauchy}
\end{figure}

In contrast, trace plots may be much less clear in certain situations. To
illustrate this point, we show trace plots of the Cauchy model in the nominal 
and alternative parameterizations side by side in Figure \ref{fig:trace-cauchy}.
Recall that the computation converged well in the alternative parameterization
but not in the nominal parameterization.

\hypertarget{half-cauchy-distribution-with-nominal-parameterization}{%
\subsubsection*{Half-Cauchy distribution with nominal
parameterization}\label{half-cauchy-distribution-with-nominal-parameterization}}

Half-Cauchy priors for non-negative parameters are common and 
often specified via the nominal parameterization.
In this example, we set independent half-Cauchy distributions on each element
of the 50-dimensional vector $x$ constrained to be positive. Probabilistic programming
frameworks usually implement positivity constraint by 
sampling in the unconstrained \(\log(x)\) space, which
changes the geometry crucially. With this transformation, all values of \Rhat\ are less than 1.01 and ESS exceeds 400 for all parameters, indicating good performance of the sampler despite using the nominal parameterization of
the Cauchy distribution. More experiments for the half-Cauchy distribution 
can be found in the online appendix.

\hypertarget{eightschools}{%
\subsection{Hierarchical model: Eight schools}\label{eightschools}}

The eight schools problem is a classic example
\citep[see Section 5.5 in][]{BDA3}, which even in its
simplicity illustrates typical problems in inference for
hierarchical models. We can parameterize this simple model
in at least two ways. The centered parameterization $(\theta, \mu,
\tau, \sigma)$ is,
\begin{align*}
\theta_j &\sim \N(\mu, \tau) \\
y_j &\sim \N(\theta_j, \sigma_j).
\end{align*}

In contrast, the non-centered parameterization
$(\tilde{\theta}, \mu, \tau, \sigma)$ can be written as,
\begin{align*}
\tilde{\theta}_j &\sim \N(0, 1) \\
\theta_j &= \mu + \tau \tilde{\theta}_j \\
y_j &\sim \N(\theta_j, \sigma_j).
\end{align*}
In both cases, $\theta_j$ are the treatment effects in the eight schools,
and $\mu, \tau$ represent the population mean and standard deviation 
of the distribution of these effects. In the centered
parameterization, the $\theta$ are parameters, whereas in the
non-centered parameterization, the $\tilde{\theta}$ are parameters and
$\theta$ is a derived quantity.

Geometrically, the centered parameterization exhibits a funnel shape
that contracts into a region of strong curvature around the population
mean when faced with small
values of the population standard deviation $\tau$, making it difficult for many simple
Markov chain methods to adequately explore the full distribution of this
parameter. In the following, we will focus on analyzing convergence of $\tau$.
The online appendix contains more detailed analysis of different 
algorithm variants and results of longer chains.

\hypertarget{a-centered-eight-schools-model}{%
\subsubsection*{A centered eight schools
model}\label{a-centered-eight-schools-model}}

Instead of the default options, we run the centered parameterization
model with more conservative settings of the HMC sample to reduce the
probability of getting divergent transitions, which bias the obtained estimates 
if they occur;  for details see \cite{StanManual.2.18.0}.
Still, we observe a lot of divergent
transitions, which in itself is already a sufficient indicator of
convergence problems. We can also use \Rhat\ and ESS
diagnostics to recognize problematic parts of the posterior. The latter
two have the advantage over the divergent transitions diagnostic that they
can be used with all MCMC algorithms not only with HMC.

\begin{figure}[tp]
  \centering
  \begin{minipage}{0.48\textwidth}
  \includegraphics[width=0.98\textwidth]{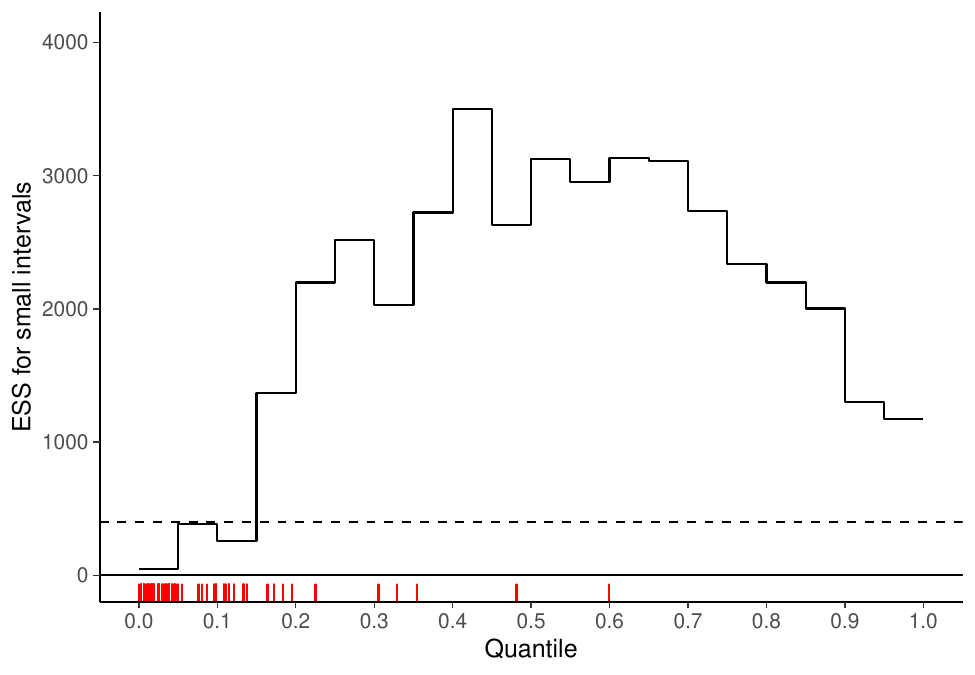}
  \caption{Local efficiency of small-interval probability estimates of $\tau$ 
  for the eight schools model with centered parameterization. The dashed line shows the
    recommended threshold of $400$. Red ticks show 
  the position of divergent transitions.}
  \label{fig:local-ess-fit-cp-1}
\end{minipage}
\hfill
  \begin{minipage}{0.48\textwidth}
  \includegraphics[width=0.98\textwidth]{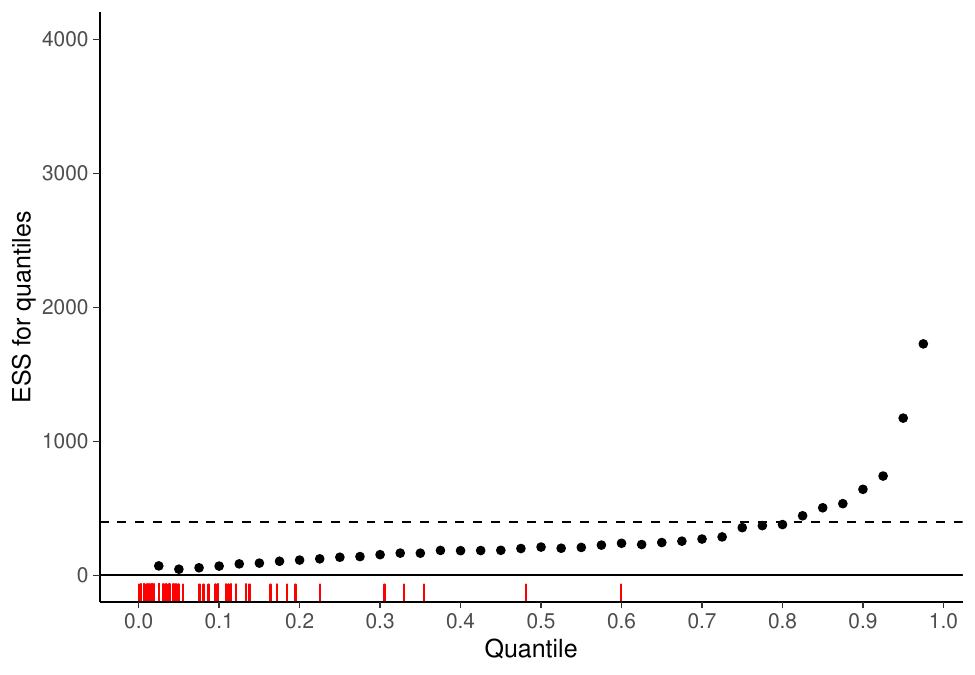}
  \caption{Efficiency of quantile estimates of $\tau$ for the eight schools
  model with centered parameterization. The dashed line shows the
    recommended threshold of $400$. Red ticks show the position of
  divergent transitions.\\~}
  \label{fig:quantile-ess-fit-cp-1}
 \end{minipage}
\end{figure}

Bulk-ESS and tail-ESS for the between-school standard deviation $\tau$
are 67 and 82, respectively. Both are much less than 400, indicating we
should investigate that parameter more carefully.
Figures~\ref{fig:local-ess-fit-cp-1} and
\ref{fig:quantile-ess-fit-cp-1} show the sampling efficiency for the
small-interval probability and quantile estimates.
The sampler has difficulties in exploring small $\tau$ values. As the
sampling efficiency for small $\tau$ values is practically zero, we
may assume that we miss substantial amount of posterior mass and
get biased estimates. In this case, the severe sampling problems for
small $\tau$ values is reflected in the sampling efficiency for all
quantiles. Red ticks, which show the position of iterations with divergences,
have concentrated to small $\tau$ values, which gives us another indication
of problems in exploring small values.

\begin{figure}[tp]
  \centering
  \begin{minipage}{0.48\textwidth}
  \includegraphics[width=0.98\textwidth]{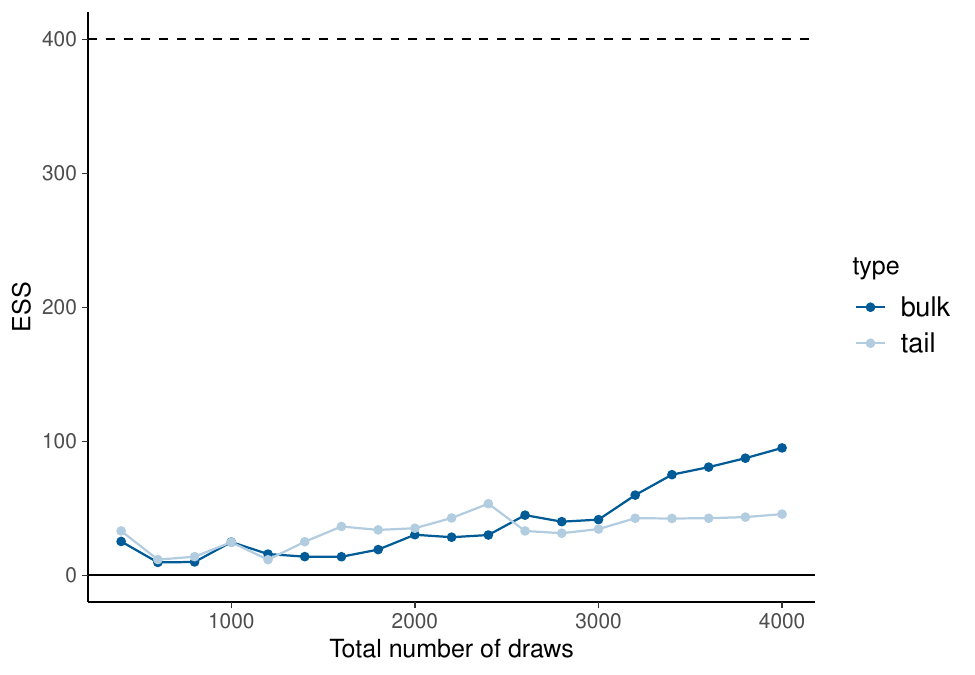}
  \caption{Estimated effective sample sizes of $\tau$ with increasing number of iterations
  for the eight schools model with centered parameterization. The dashed line shows the
    recommended threshold of $400$.}
  \label{fig:change-ess-fit-cp-1}
\end{minipage}
\hfill
  \begin{minipage}{0.48\textwidth}
  \includegraphics[width=0.98\textwidth]{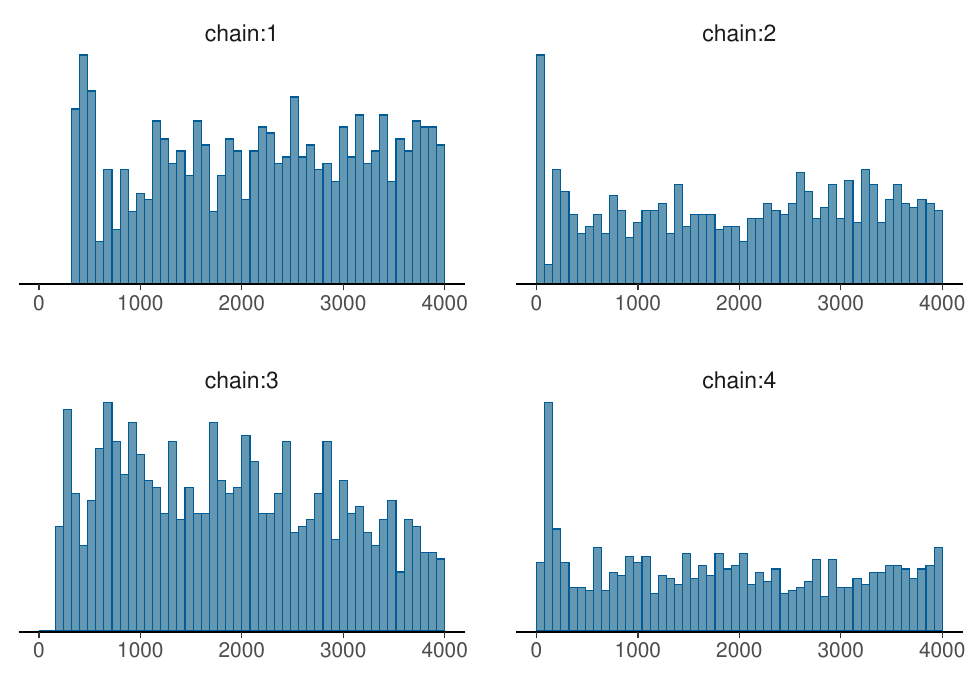}
  \caption{Rank plots of posterior draws of $\tau$ from four chains for 
  the eight schools model with centered parameterization.}
  \label{fig:hist-fit-cp-1}
\end{minipage}
\end{figure}

Figure~\ref{fig:change-ess-fit-cp-1} shows how the estimated effective
sample sizes change when we use more and more draws. Here we do not
see sudden changes, but both bulk-ESS and tail-ESS are consistently low. 
In line with the other findings, rank plots of $\tau$ displayed in
Figure~\ref{fig:hist-fit-cp-1} clearly show problems in the mixing of
the chains.  In particular, the rank plot for the first chain indicates that it was unable to 
explore the lower-end of the posterior range, while the spike in the rank plot 
for chain 2 indicates that it spent too much time stuck in these values.
More experiments can be found in Appendices~C and D as well as in the online appendix.

\hypertarget{non-centered-eight-schools-model}{%
\subsubsection*{Non-centered eight schools
model}\label{non-centered-eight-schools-model}}

\begin{figure}[tp]
  \begin{minipage}{0.48\textwidth}
  \centering
  \includegraphics[width=0.98\textwidth]{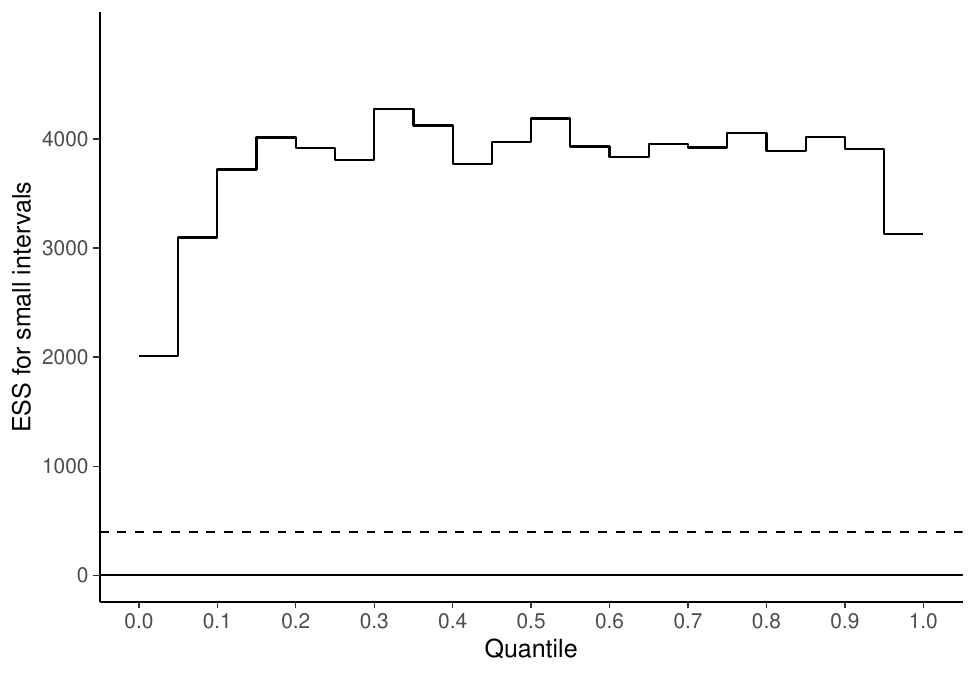}
  \caption{Local efficiency of small-interval probability estimates of $\tau$ 
  for the  eight schools model with the non-centered parameterization. The dashed line shows the
    recommended threshold of $400$.}
  \label{fig:local-ess-fit-ncp2-1}
\end{minipage}
\hfill
  \begin{minipage}{0.48\textwidth}
  \includegraphics[width=0.98\textwidth]{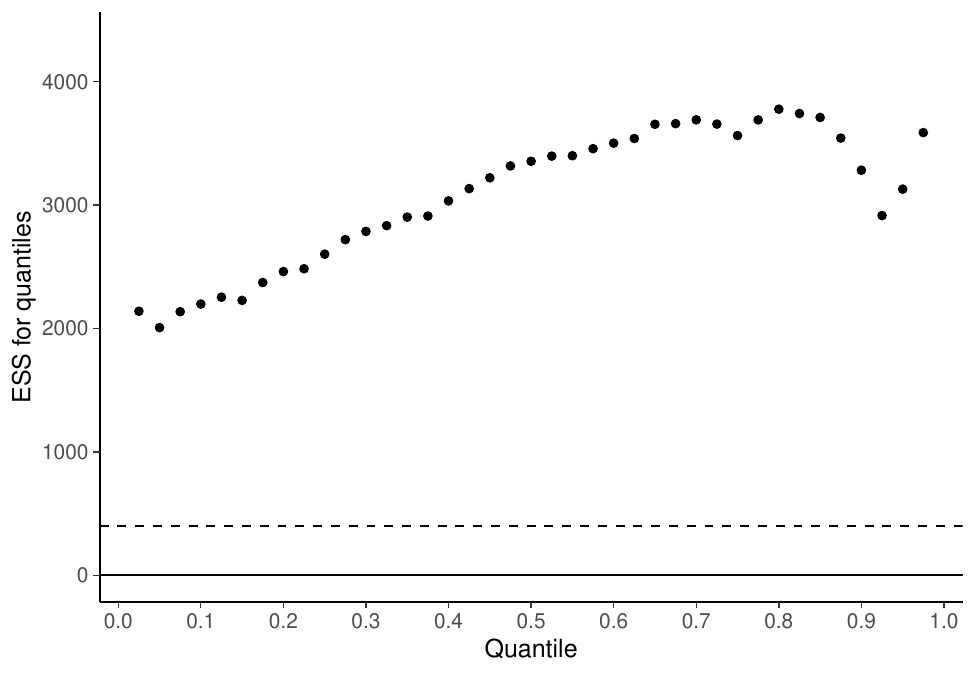}
  \caption{Efficiency of quantile estimates of $\tau$ for the eight schools 
  model with the non-centered parameterization. The dashed line shows the
    recommended threshold of $400$.}
  \label{fig:quantile-ess-fit-ncp2-1}
\end{minipage}
\end{figure}

\begin{figure}[tp]
  \centering
  \includegraphics[width=0.47\textwidth]{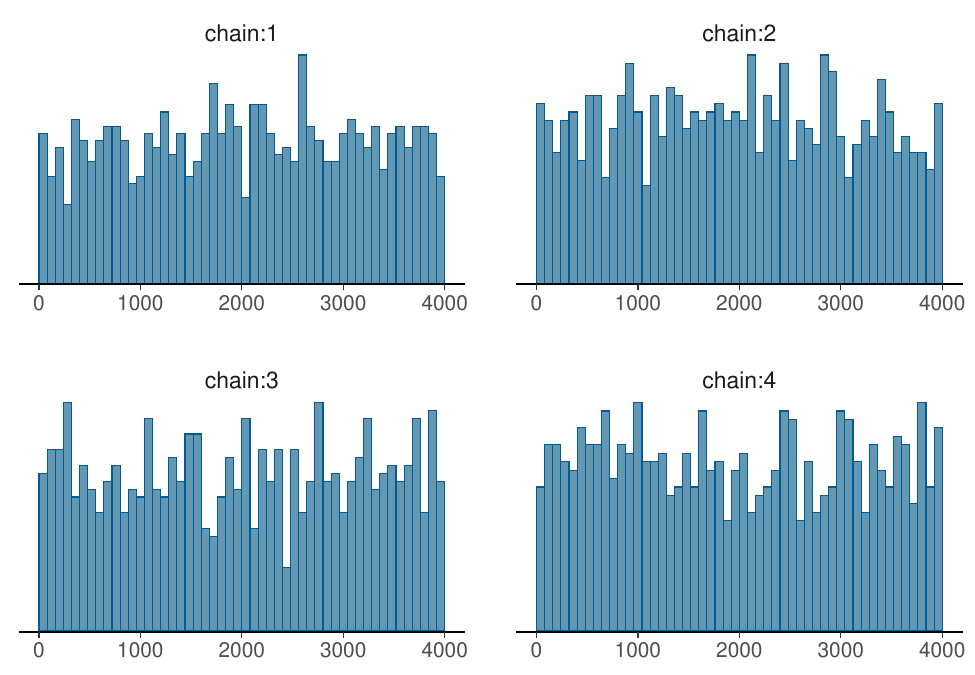}
  \caption{Rank plots of posterior draws of $\tau$ from four chains for 
  the eight schools model with non-centered parameterization.}
  \label{fig:hist-fit-ncp2-1}
\end{figure}

For hierarchical models, the corresponding non-centered
parameterization often works better \citep{Betancourt+Girolami:2019}. 
For reasons of comparability, we use the 
same conservative sampler settings as for the centered parameterization model. 
For the non-centered parameterization, we do not observe divergences or 
other warnings.
All values of \Rhat\ are less than 1.01 and ESS exceeds 400, indicating a much
better efficiency of the non-centered parameterization.
Figures~\ref{fig:local-ess-fit-ncp2-1} and
\ref{fig:quantile-ess-fit-ncp2-1} show the efficiency of small-interval
probability estimates and the efficiency of quantile estimates for
$\tau$.
Small $\tau$ values are still more difficult to explore, but the relative 
efficiency is good. The rank plots of $\tau$ Figure~\ref{fig:hist-fit-ncp2-1} 
show no substantial differences between chains.

\hypertarget{refs}{}

\bibliography{rhat}

\begin{thebibliography}{39}
\providecommand{\natexlab}[1]{#1}
\providecommand{\url}[1]{\texttt{#1}}
\expandafter\ifx\csname urlstyle\endcsname\relax
  \providecommand{\doi}[1]{doi: #1}\else
  \providecommand{\doi}{doi: \begingroup \urlstyle{rm}\Url}\fi

\bibitem[Betancourt(2017)]{betancourt2017conceptual}
Michael Betancourt.
\newblock A conceptual introduction to {Hamiltonian Monte Carlo}.
\newblock \emph{arXiv preprint arXiv:1701.02434}, 2017.

\bibitem[Betancourt and Girolami(2019)]{Betancourt+Girolami:2019}
Michael Betancourt and Mark Girolami.
\newblock {Hamiltonian} {Monte} {Carlo} for hierarchical models.
\newblock In \emph{Current Trends in Bayesian Methodology with Applications},
  pages 79--101. Chapman and Hall/CRC, 2019.

\bibitem[Blom(1958)]{Blom:1958}
Gunnar Blom.
\newblock \emph{Statistical Estimates and Transformed Beta-Variables}.
\newblock Wiley; New York, 1958.

\bibitem[Brooks and Gelman(1998)]{Brooks+Gelman:1998}
Stephen~P. Brooks and Andrew Gelman.
\newblock General methods for monitoring convergence of iterative simulations.
\newblock \emph{Journal of Computational and Graphical Statistics}, 7\penalty0
  (4):\penalty0 434--455, 1998.

\bibitem[Carpenter et~al.(2017)Carpenter, Gelman, Hoffman, Lee, Goodrich,
  Betancourt, Brubaker, Guo, Li, and Riddell]{Stan:JSS:2017}
Bob Carpenter, Andrew Gelman, Matthew Hoffman, Daniel Lee, Ben Goodrich,
  Michael Betancourt, Marcus Brubaker, Jiqiang Guo, Peter Li, and Allen
  Riddell.
\newblock Stan: A probabilistic programming language.
\newblock \emph{Journal of Statistical Software, Articles}, 76\penalty0
  (1):\penalty0 1--32, 2017.
\newblock \doi{10.18637/jss.v076.i01}.

\bibitem[Chernoff and Savage(1958)]{Chernoff+Savage:1958}
Herman Chernoff and I.~Richard Savage.
\newblock Asymptotic normality and efficiency of certain nonparametric test
  statistics.
\newblock \emph{Annals of Mathematical Statistics}, 29\penalty0 (4):\penalty0
  972--994, 1958.

\bibitem[Cowles and Carlin(1996)]{Cowles+Carlin:1996}
Mary~Kathryn Cowles and Bradley~P. Carlin.
\newblock {Markov} chain {Monte} {Carlo} convergence diagnostics: A comparative
  review.
\newblock \emph{Journal of the American Statistical Association}, 91\penalty0
  (434):\penalty0 883--904, 1996.

\bibitem[de~Valpine et~al.(2017)de~Valpine, Turek, Paciorek, Anderson-Bergman,
  Lang, and Bodik]{nimble}
Perry de~Valpine, Daniel Turek, Christopher~J. Paciorek, Clifford
  Anderson-Bergman, Duncan~Temple Lang, and Rastislav Bodik.
\newblock Programming with models: Writing statistical algorithms for general
  model structures with {NIMBLE}.
\newblock \emph{Journal of Computational and Graphical Statistics}, 26\penalty0
  (2):\penalty0 403--413, 2017.

\bibitem[Doss et~al.(2014)Doss, Flegal, Jones, and
  Neath]{Doss+etal:2014:MCMC-quantiles}
Charles~R. Doss, James~M. Flegal, Galin~L. Jones, and Ronald~C. Neath.
\newblock {Markov} chain {Monte} {Carlo} estimation of quantiles.
\newblock \emph{Electronic Journal of Statistics}, 8\penalty0 (2):\penalty0
  2448--2478, 2014.

\bibitem[Fisher and Yates(1938)]{Fisher+Yates:1938}
Ronald~A. Fisher and Frank Yates.
\newblock \emph{Statistical Tables for Biological, Agricultural, and Medical
  Research}.
\newblock Oliver \& Boyd; Edinburgh, 1938.

\bibitem[Flegal and Jones(2010)]{Flegal+Jones:2010}
James~M. Flegal and Galin~L. Jones.
\newblock Batch means and spectral variance estimators in {Markov} chain
  {Monte} {Carlo}.
\newblock \emph{Annals of Statistics}, 38\penalty0 (2):\penalty0 1034--1070,
  2010.

\bibitem[Friedman(1937)]{Friedman:1937}
Milton Friedman.
\newblock The use of ranks to avoid the assumption of normality implicit in the
  analysis of variance.
\newblock \emph{Journal of the American Statistical Association}, 32\penalty0
  (200):\penalty0 675--701, 1937.

\bibitem[Gelman and Rubin(1992)]{Gelman+Rubin:1992}
Andrew Gelman and Donald~B. Rubin.
\newblock Inference from iterative simulation using multiple sequences (with
  discussion).
\newblock \emph{Statistical Science}, 7\penalty0 (4):\penalty0 457--511, 1992.

\bibitem[Gelman et~al.(2003)Gelman, Carlin, Stern, and Rubin]{BDA2}
Andrew Gelman, John~B. Carlin, Hal~S. Stern, and Donald~R. Rubin.
\newblock \emph{Bayesian Data Analysis, second edition}.
\newblock Chapman \& Hall, 2003.

\bibitem[Gelman et~al.(2008)Gelman, Huang, van Dyk, and
  Boscardin]{Gelman+Huang+vanDyk+Boscardin:2008}
Andrew Gelman, Zaiying Huang, David van Dyk, and W.~John Boscardin.
\newblock Using redundant parameters to fit hierarchical models.
\newblock \emph{Journal of Computational and Graphical Statistics},
  17:\penalty0 95--122, 2008.

\bibitem[Gelman et~al.(2013)Gelman, Carlin, Stern, Dunson, Vehtari, and
  Rubin]{BDA3}
Andrew Gelman, John~B. Carlin, Hal~S. Stern, David~B. Dunson, Aki Vehtari, and
  Donald~B. Rubin.
\newblock \emph{Bayesian Data Analysis, third edition}.
\newblock CRC Press, 2013.

\bibitem[Geyer(1992)]{Geyer:1992}
Charles~J. Geyer.
\newblock Practical {Markov} chain {Monte} {Carlo}.
\newblock \emph{Statistical Science}, 7:\penalty0 473--483, 1992.

\bibitem[Geyer(2011)]{Geyer:2011}
Charles~J. Geyer.
\newblock Introduction to {Markov} chain {Monte} {Carlo}.
\newblock In S.~Brooks, A.~Gelman, G.~L. Jones, and X.~L. Meng, editors,
  \emph{Handbook of Markov Chain Monte Carlo}. CRC Press, 2011.

\bibitem[Hastings(1970)]{Hastings:1970}
W.~K. Hastings.
\newblock {Monte} {Carlo} sampling methods using {Markov} chains and their
  applications.
\newblock \emph{Biometrika}, 57\penalty0 (1):\penalty0 97--109, 1970.

\bibitem[Hoffman and Gelman(2014)]{Hoffman+Gelman:2014}
Matthew~D. Hoffman and Andrew Gelman.
\newblock The {No-U-Turn} {Sampler}: Adaptively setting path lengths in
  {Hamiltonian} {Monte} {Carlo}.
\newblock \emph{Journal of Machine Learning Research}, 15:\penalty0 1593--1623,
  2014.
\newblock URL \url{http://jmlr.org/papers/v15/hoffman14a.html}.

\bibitem[Jacob et~al.(2017)Jacob, O'Leary, and Atchad{\'e}]{jacob2017unbiased}
Pierre~E. Jacob, John O'Leary, and Yves~F. Atchad{\'e}.
\newblock Unbiased {Markov} chain {Monte} {Carlo} with couplings.
\newblock \emph{arXiv preprint arXiv:1708.03625}, 2017.

\bibitem[Julier and Uhlmann(1997)]{Julier+Uhlman:1997:unscented}
Simon~J Julier and Jeffrey~K Uhlmann.
\newblock New extension of the kalman filter to nonlinear systems.
\newblock In \emph{Proc. SPIE 3068, Signal processing, sensor fusion, and
  target recognition VI}, pages 182--193. SPIE, 1997.

\bibitem[Kong et~al.(1994)Kong, Liu, and Wong]{Kong+Liu+Wong:1994}
Augustine Kong, Jun~S. Liu, and Wing~Hung Wong.
\newblock Sequential imputations and {Bayesian} missing data problems.
\newblock \emph{Journal of the American Statistical Association}, 89\penalty0
  (425):\penalty0 278--288, 1994.

\bibitem[Laurmann and Gates(1977)]{Laurmann+Gates:1977}
John~A. Laurmann and W.~Lawrence Gates.
\newblock Statistical considerations in the evaluation of climatic experiments
  with atmospheric general circulation models.
\newblock \emph{Journal of the Atmospheric Sciences}, 34\penalty0 (8):\penalty0
  1187--1199, 1977.

\bibitem[Liu et~al.(2016)Liu, Nordman, and
  Meeker]{Liu+etal:2016:MCMC-quantiles}
Jia Liu, Daniel~J. Nordman, and William~Q. Meeker.
\newblock The number of {MCMC} draws needed to compute {Bayesian} credible
  bounds.
\newblock \emph{The American Statistician}, 70\penalty0 (3):\penalty0 275--284,
  2016.

\bibitem[Lunn et~al.(2009)Lunn, Spiegelhalter, Thomas, and
  Best]{BUGSproject:2009}
David Lunn, David Spiegelhalter, Andrew Thomas, and Nicky Best.
\newblock The {BUGS} project: Evolution, critique and future directions.
\newblock \emph{Statistics in Medicine}, 28\penalty0 (25):\penalty0 3049--3067,
  2009.

\bibitem[Lunn et~al.(2000)Lunn, Thomas, Best, and Spiegelhalter]{WinBUGS:2000}
David~J Lunn, Andrew Thomas, Nicky Best, and David Spiegelhalter.
\newblock {WinBUGS}---a {Bayesian} modelling framework: Concepts, structure,
  and extensibility.
\newblock \emph{Statistics and Computing}, 10\penalty0 (4):\penalty0 325--337,
  2000.

\bibitem[Mengersen et~al.(1999)Mengersen, Robert, and
  Guihenneuc-Jouyaux]{Mengersen+etal:1999}
Kerrie~L. Mengersen, Christian~P. Robert, and Chantal Guihenneuc-Jouyaux.
\newblock {MCMC} convergence diagnostics: A review.
\newblock In Jose~M. Bernardo, James~O. Berger, and A.~P. Dawid, editors,
  \emph{Bayesian Statistics 6}, pages 415--440. Oxford University Press, 1999.

\bibitem[Neal(2003)]{Neal:2003}
Radford~M. Neal.
\newblock Slice sampling.
\newblock \emph{Annals of Statistics}, 31\penalty0 (3):\penalty0 705--767,
  2003.

\bibitem[Plummer(2003)]{plummer2003jags}
Martyn Plummer.
\newblock {JAGS}: A program for analysis of {Bayesian} graphical models using
  {Gibbs} sampling.
\newblock In \emph{Proceedings of the 3rd International Workshop on Distributed
  Statistical Computing}, volume 124, 2003.

\bibitem[Plummer et~al.(2006)Plummer, Best, Cowles, and Vines]{coda2006}
Martyn Plummer, Nicky Best, Kate Cowles, and Karen Vines.
\newblock {CODA}: Convergence diagnosis and output analysis for {MCMC}.
\newblock \emph{R News}, 6\penalty0 (1):\penalty0 7--11, 2006.
\newblock URL \url{https://journal.r-project.org/archive/}.

\bibitem[Raftery and Lewis(1992)]{Raftery+Lewis:1992a}
Adrian~E. Raftery and Steven~M. Lewis.
\newblock How many iterations in the {Gibbs} sampler?
\newblock In J.~M. Bernardo, J.~O. Berger, A.~P. Dawid, and A.~F.~M. Smith,
  editors, \emph{Bayesian Statistics 4}, pages 763--773. Oxford University
  Press, 1992.

\bibitem[Robert and Casella(2004)]{Robert+Casella:2004}
Christian~P. Robert and George Casella.
\newblock \emph{Monte Carlo Statistical Methods}.
\newblock Springer, second edition, 2004.

\bibitem[Salvatier et~al.(2016)Salvatier, Wiecki, and Fonnesbeck]{pymc3}
John Salvatier, Thomas~V. Wiecki, and Christopher Fonnesbeck.
\newblock Probabilistic programming in {Python} using {PyMC3}.
\newblock \emph{PeerJ Computer Science}, 2:\penalty0 e55, 2016.

\bibitem[Sorensen et~al.(1995)Sorensen, Andersen, Gianola, and
  Korsgaard]{Sorensen+etal:1995}
D.~A. Sorensen, S.~Andersen, D.~Gianola, and I.~Korsgaard.
\newblock Bayesian inference in threshold models using {Gibbs} sampling.
\newblock \emph{Genetics Selection Evolution}, 27\penalty0 (3):\penalty0 229,
  1995.

\bibitem[{Stan Development Team}(2018{\natexlab{a}})]{RStanARM.2.17}
{Stan Development Team}.
\newblock {RStanArm}: {Bayesian} applied regression modeling via {Stan}. {R}
  package version 2.17.4, 2018{\natexlab{a}}.
\newblock URL \url{http://mc-stan.org}.

\bibitem[{Stan Development Team}(2018{\natexlab{b}})]{StanManual.2.18.0}
{Stan Development Team}.
\newblock {Stan Modeling Language Users Guide and Reference Manual}. version
  2.18.0, 2018{\natexlab{b}}.
\newblock URL \url{http://mc-stan.org}.

\bibitem[Vats and Knudson(2018)]{vats2018revisiting}
Dootika Vats and Christina Knudson.
\newblock Revisiting the {Gelman}-{Rubin} diagnostic.
\newblock \emph{arXiv preprint arXiv:1812.09384}, 2018.

\bibitem[Wellek(2010)]{Wellek:2010:testing}
Stefan Wellek.
\newblock \emph{Testing Statistical Hypotheses of Equivalence and
  Noninferiority}.
\newblock Chapman and Hall/CRC, 2010.

\end{thebibliography}

\newpage

\hypertarget{AppendixD}{%
\subsection*{Appendix A: Normal distributions with additional trend,
shift, or scaling}\label{AppendixA}}
\addcontentsline{toc}{subsection}{Appendix A: Normal distributions with
additional trend, shift or scaling}

Here we demonstrate the behavior of non-split-\(\widehat{R}\),
split-\(\widehat{R}\), and bulk-ESS to detect various simulated cases
presenting non-convergence behavior. We generate four varying length
chains of iid normally distributed values, and then modify them to
simulate three convergence problems:
\begin{itemize}
\tightlist
\item All chains have the same trend and a similar marginal
  distribution. This can happen in case of slow mixing and all chains
  initialized near each other far from the typical set.
\item One of the chains has a different mean. This can happen in case
  of slow mixing, weak identifiability of one or several parameters, or
  multimodality.
\item One of the chains having a lower marginal variance. This can
  happen in case of slow mixing, multimodality, or one of the chains
  having different mixing efficiency.
\end{itemize}

The code for these simulations can be found in the online appendix.

\begin{figure}[tp]
  \centering
  \begin{minipage}{0.48\textwidth}
  \includegraphics[width=0.98\textwidth]{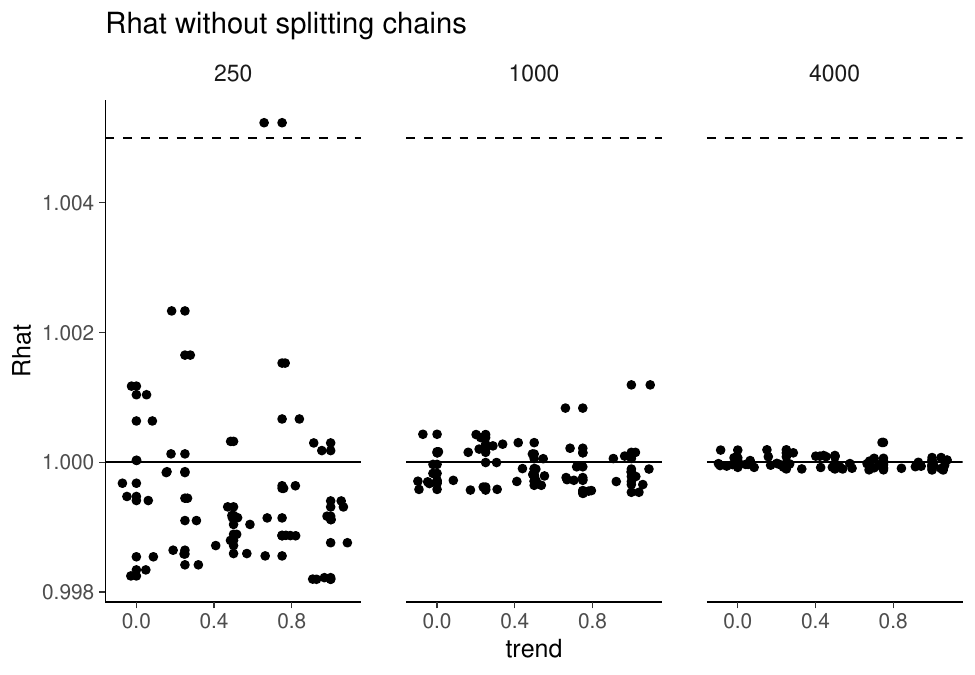}
  \caption{\(\widehat{R}\) without splitting for varying chain lengths
    for chains which have the same linear trend and a 
    similar marginal distribution.}
  \label{fig:rhat-same-trend-1}
\end{minipage}
\hfill
  \begin{minipage}{0.48\textwidth}
  \includegraphics[width=0.98\textwidth]{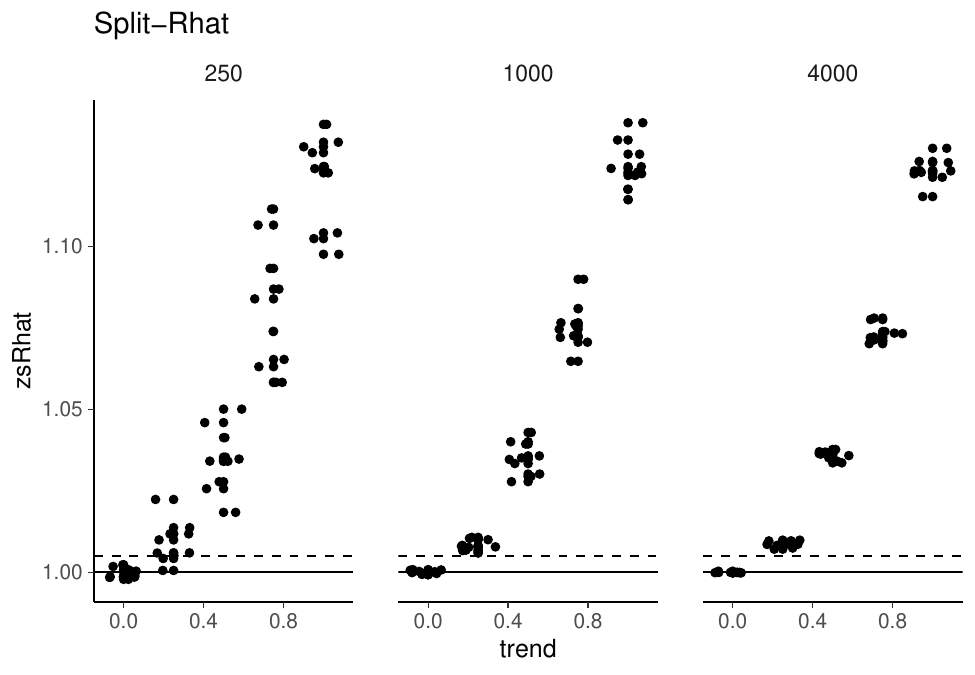}
  \caption{Split-\(\widehat{R}\) for varying chain lengths
    for chains which have the same linear trend and a similar marginal
    distribution.}
  \label{fig:zsrhat-same-trend-1}
\end{minipage}
\end{figure}

\hypertarget{adding-the-same-trend-to-all-chains}{%
\paragraph{All chains have the same trend.}\label{adding-the-same-trend-to-all-chains}}
First, we draw all the chains from the same $\N(0, 1)$ distribution plus
a linear trend (i.e., $\theta^{(s)} = e^{(s)} + c s$ where $e^{(s)}$ is $\N(0, 1)$ 
distributed, $s$ is the iteration indicator, and $c$ is the strength of the 
linear trend). Figure~\ref{fig:rhat-same-trend-1}
shows that if we don't split chains, \(\widehat{R}\) misses the trends
if all chains still have a similar marginal distribution.
Figure~\ref{fig:zsrhat-same-trend-1} shows that split-\(\widehat{R}\)
detects the trend, even if the marginals of the chains are similar. If
we use a threshold of \(1.01\), we can detect trends which account for
2\% or more of the total marginal variance. If we use a threshold of
\(1.1\), we detect trends which account for 30\% or more of the total
marginal variance.

\begin{figure}[tp]
  \centering
  \begin{minipage}{0.48\textwidth}
  \includegraphics[width=0.98\textwidth]{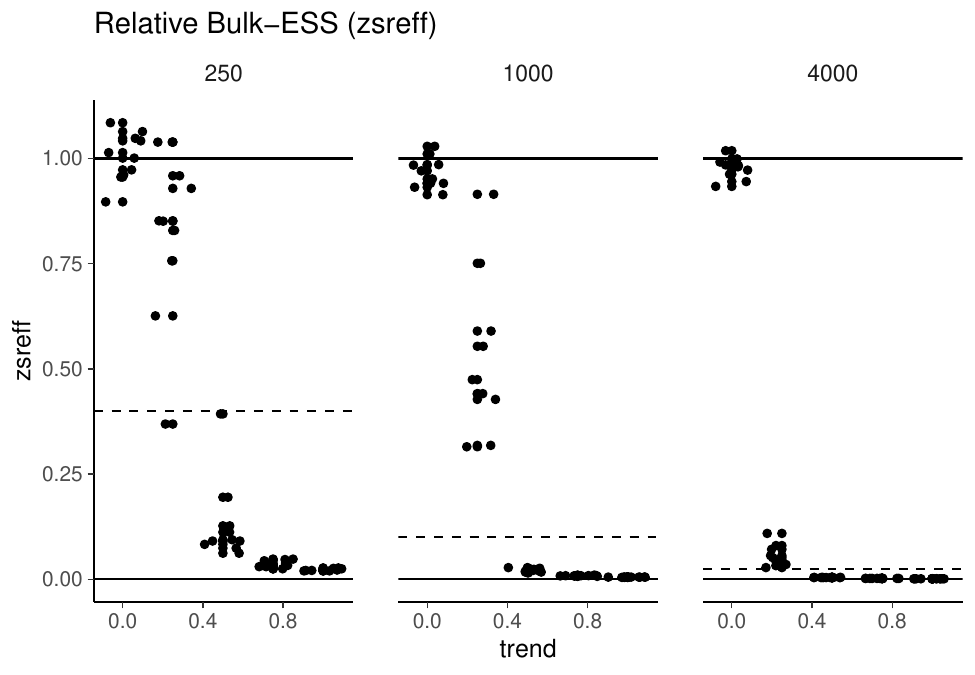}
  \caption{Relative bulk-ESS for varying chain lengths for chains which have
    the same trend and a similar marginal distribution. The dashed
    lines indicate the threshold \(S_{\rm eff} > 400\) at which we
    would consider the effective sample size to be sufficient.}
  \label{fig:zsreff-same-trend-1}
\end{minipage}
\hfill
  \begin{minipage}{0.48\textwidth}
  \includegraphics[width=0.98\textwidth]{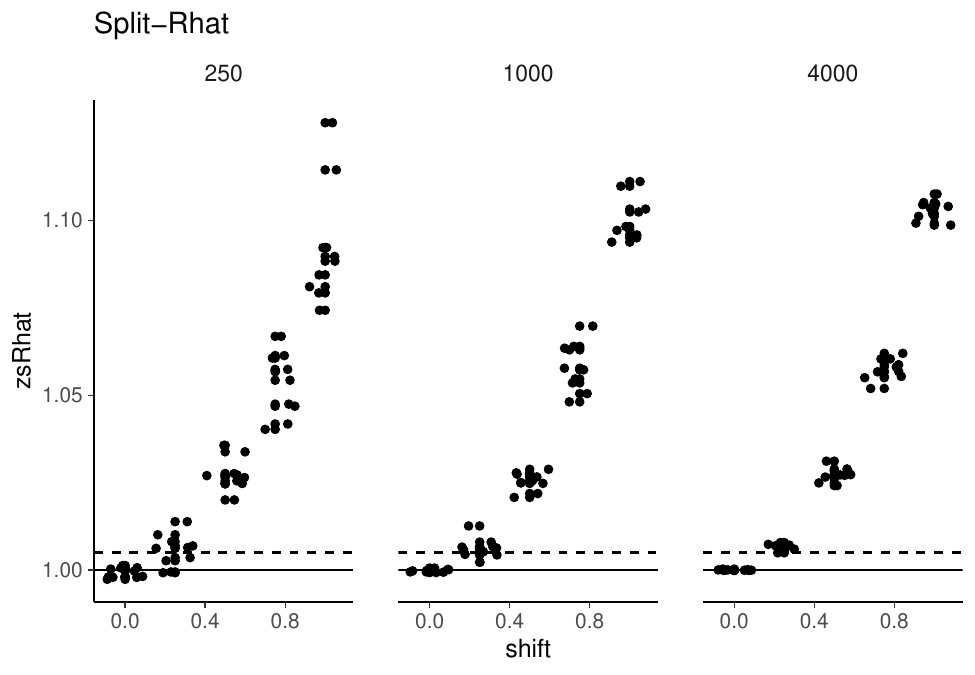}
  \caption{Split-\(\widehat{R}\) for varying chain lengths
    for chains with one sampled with a different mean than the others.\\~\\~}
  \label{fig:zsrhat-shifted-chain-1}
\end{minipage}
\end{figure}

The effective sample size is based on split-\(\widehat{R}\) and
within-chain autocorrelation. Figure~\ref{fig:zsreff-same-trend-1}
shows the relative bulk-ESS divided by \(S\) for easier
comparison between different values of \(S\).
Split-\(\widehat{R}\) is more sensitive to trends for
small sample sizes, but ESS becomes more sensitive for larger sample
sizes (as autocorrelations can be estimated more accurately).

\begin{figure}[tp]
  \centering
  \begin{minipage}{0.48\textwidth}
  \includegraphics[width=0.98\textwidth]{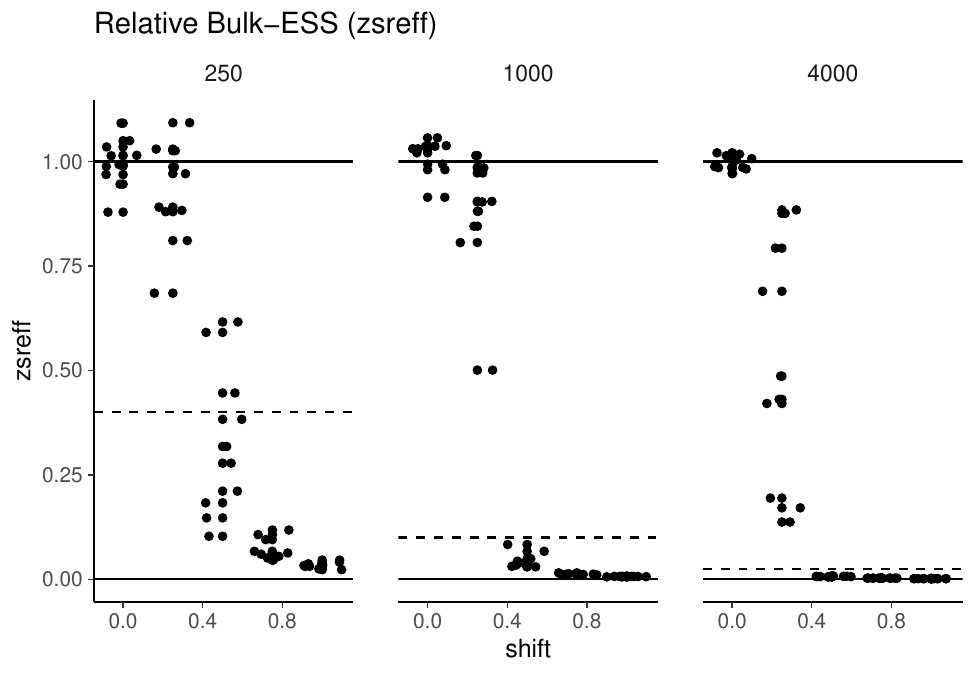}
  \caption{Relative bulk-ESS for varying chain lengths for chains with one
    sampled with a different mean than the others. The dashed lines
    indicate the threshold \(S_{\rm eff} > 400\) at which we would
    consider the effective sample size to be sufficient.}
  \label{fig:zsreff-shifted-chain-1}
\end{minipage}
\hfill
  \begin{minipage}{0.48\textwidth}
  \includegraphics[width=0.98\textwidth]{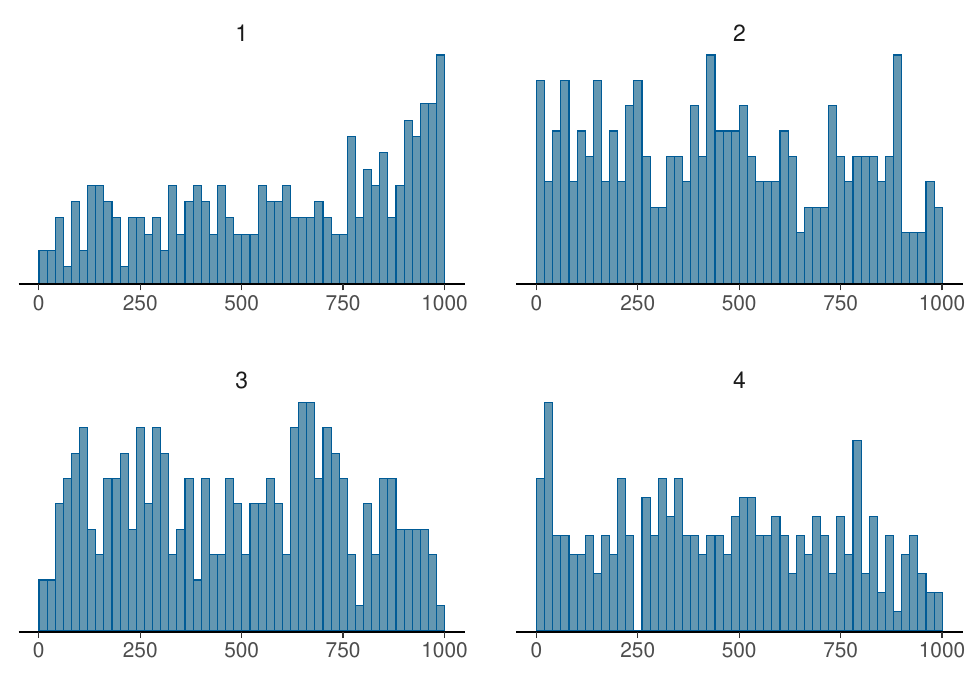}
  \caption{Rank plots of posterior draws from four chains with one
    sampled with a different mean than the others.\\~\\}
  \label{fig:hist-shifted-chain-1}
\end{minipage}
\end{figure}

\hypertarget{shifting-one-chain}{%
\paragraph{Shifting one chain.}\label{shifting-one-chain}}
Second, we draw all the chains from the same $\N(0, 1)$ distribution,
except one that is sampled with nonzero
mean. Figure~\ref{fig:zsrhat-shifted-chain-1} shows that if we use a
threshold of \(1.01\), split-\(\widehat{R}\) can detect shifts with a
magnitude of one third or more of the marginal standard deviation. If
we use a threshold of \(1.1\), split-\(\widehat{R}\) detects shifts
with a magnitude equal to or larger than the marginal standard
deviation.
Figure~\ref{fig:zsreff-shifted-chain-1} shows the the relative
bulk-ESS for the same case. The effective
sample size is not as sensitive as split-\(\widehat{R}\), but a shift
with a magnitude of half the marginal standard deviation or more will
lead to low relative efficiency when the total number of draws
increases.
Rank plots are practical way to visualize differences between
chains. Figure~\ref{fig:hist-shifted-chain-1} shows rank plots for the
case of 4 chains, 250 draws per chain, and one chain sampled with mean
0.5 instead of 0. In this case split-\(\widehat{R} = 1.05\), but the
rank plots clearly show that the first chain behaves differently.

\begin{figure}[tp]
  \centering
  \begin{minipage}{0.48\textwidth}
  \includegraphics[width=0.98\textwidth]{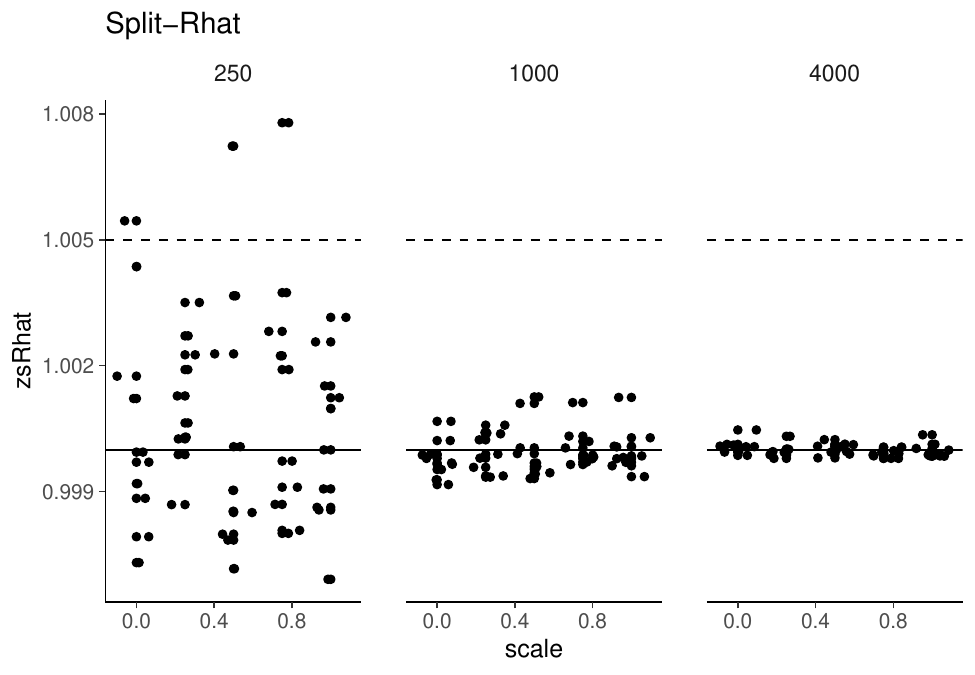}
  \caption{Split-\(\widehat{R}\) for varying chain lengths
    for chains with one sampled with a different variance than the others.}
  \label{fig:zsrhat-scaled-chain-1}
\end{minipage}
\hfill
  \begin{minipage}{0.48\textwidth}
  \includegraphics[width=0.98\textwidth]{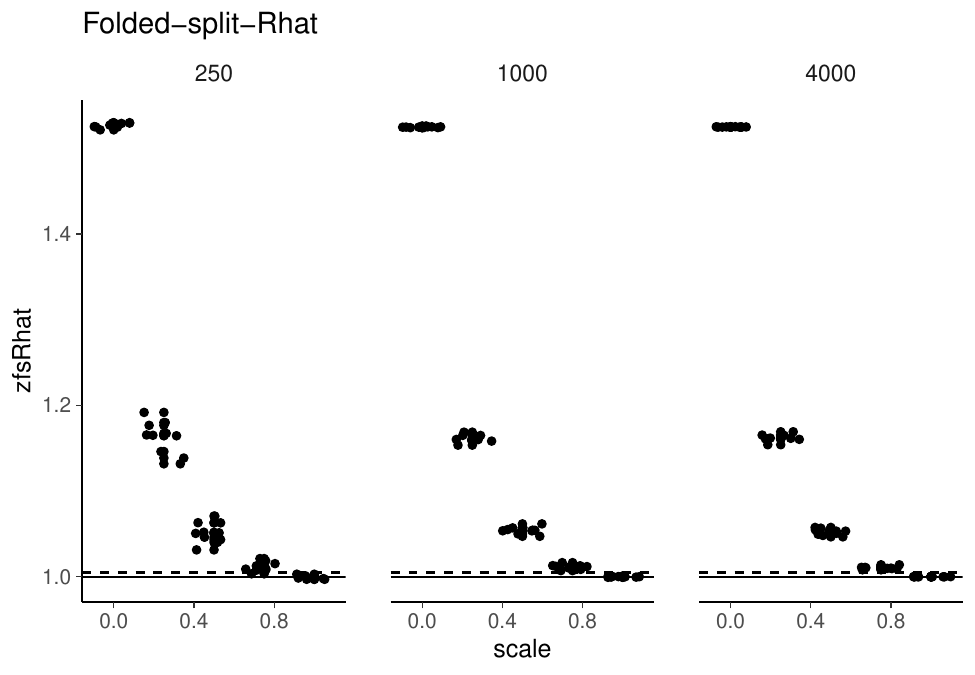}
  \caption{Folded-split-\(\widehat{R}\) for varying chain lengths
    for chains with one sampled with a different variance than the others.}
  \label{fig:zfsrhat-scaled-chain-1}
\end{minipage}
\end{figure}

\hypertarget{scaling-one-chain}{%
\paragraph{Scaling one chain.}\label{scaling-one-chain}}
For our third simulation, all the chains are from the same $\N(0, 1)$ distribution,
except one of the chains is sampled with variance less than 1.
Figure~\ref{fig:zsrhat-scaled-chain-1} shows that
split-\(\widehat{R}\) is not able to detect scale differences between
chains.
Figure~\ref{fig:zfsrhat-scaled-chain-1} shows that
folded-split-\(\widehat{R}\) which focuses on scales detects scale
differences. With a threshold of \(1.01\),
folded-split-\(\widehat{R}\) detects a chain with scale less than
\(3/4\) of the standard deviation of the others. With a threshold of
\(1.1\), folded-split-\(\widehat{R}\) detects a chain with standard
deviation less than \(1/4\) of the standard deviation of the others.

\begin{figure}[tp]
  \centering
  \begin{minipage}{0.48\textwidth}
  \includegraphics[width=0.98\textwidth]{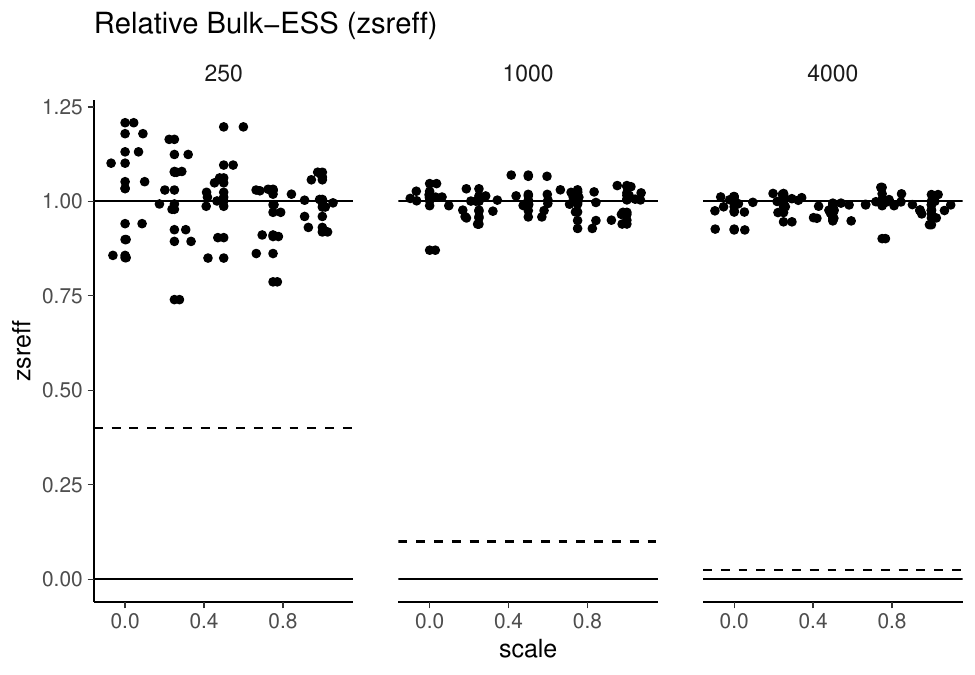}
  \caption{Relative bulk-ESS for varying chain lengths for chains with
    one sampled with a different variance than the others.}
  \label{fig:zsreff-scaled-chain-1}
\end{minipage}
\hfill
  \begin{minipage}{0.48\textwidth}
  \includegraphics[width=0.98\textwidth]{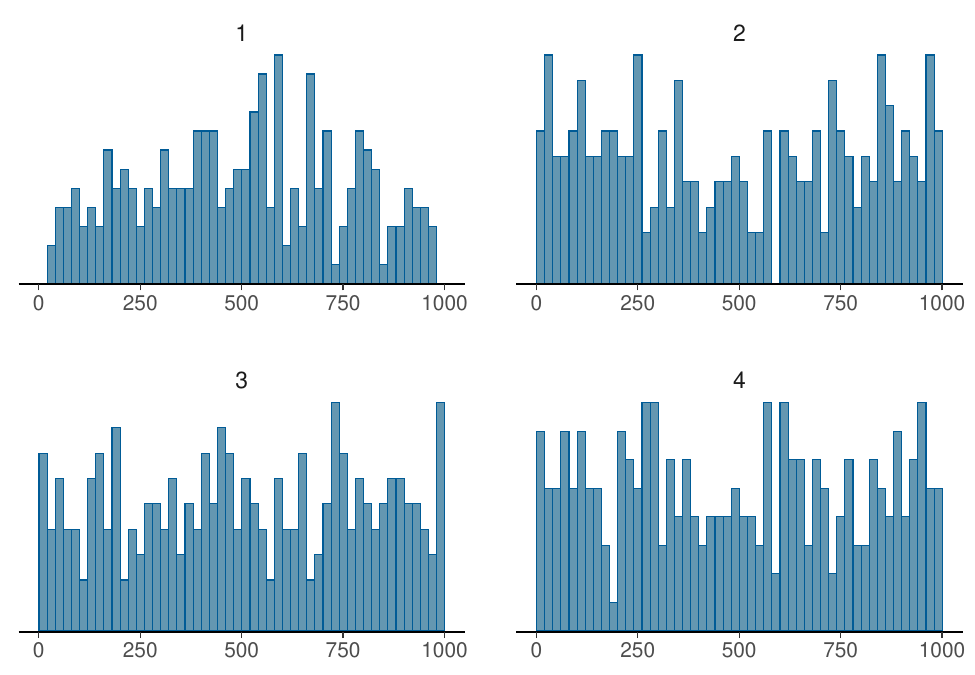}
  \caption{Rank plots of posterior draws from four chains with
    one sampled with a different variance than the others.}
  \label{fig:hist-scaled-chain-1}
\end{minipage}
\end{figure}

Figure~\ref{fig:zsreff-scaled-chain-1} shows the the relative bulk-ESS
for the same case. The bulk effective sample
size of the mean does not see a problem as it focuses on location
differences between chains.
Figure~\ref{fig:hist-scaled-chain-1} shows rank plots for the case of
4 chains, 250 draws per chain, and one chain sampled with standard
deviation 0.75 instead of 1. Although
folded-split-\(\widehat{R} = 1.06\), the rank plots clearly show that
the first chain behaves differently.

\hypertarget{AppendixB}{%
\subsection*{Appendix B: More experiments with the Cauchy distribution}\label{AppendixB}}
\addcontentsline{toc}{subsection}{Appendix B: More experiments with the Cauchy distribution}

Here we provide some additional results for the the nominal Cauchy
model presented in the main text. Instead of the default options we
increase \texttt{max\_treedepth} to \(20\), which improves the
exploration in long tails. The online appendix has additional results
for the default option case and for longer chains.

\begin{figure}[tp]
  \centering
  \includegraphics[width=0.47\textwidth]{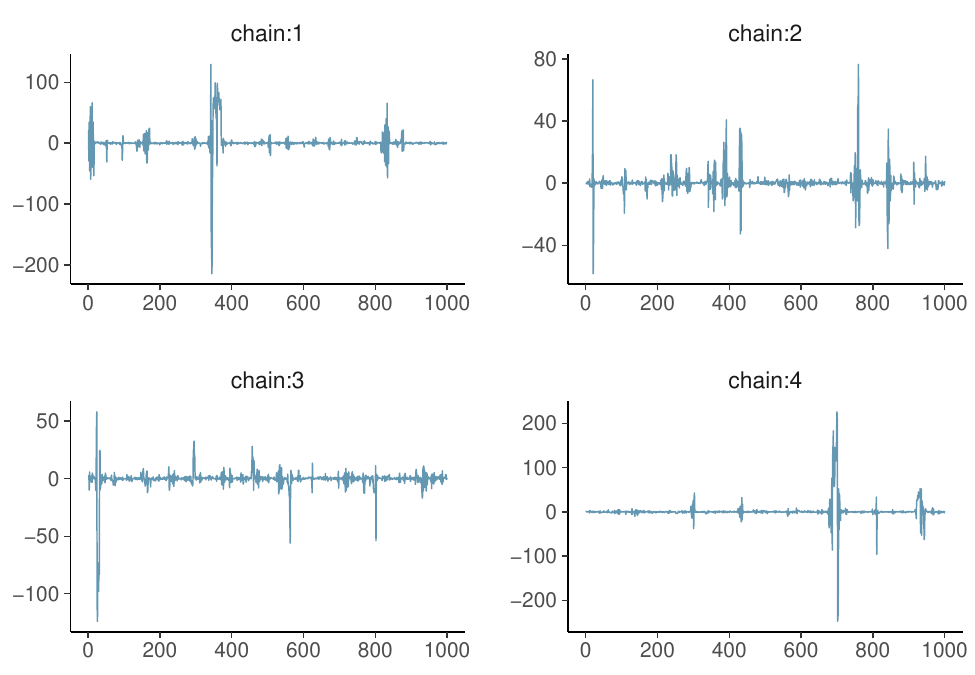}
  \caption{Trace plots of four chains for Cauchy model with nominal parameterization and \texttt{max\_treedepth}=20.\\~}
  \label{fig:trace-fit-nom-td20-1}
\end{figure}

\begin{figure}[tp]
  \centering
  \begin{minipage}{0.48\textwidth}
  \includegraphics[width=0.98\textwidth]{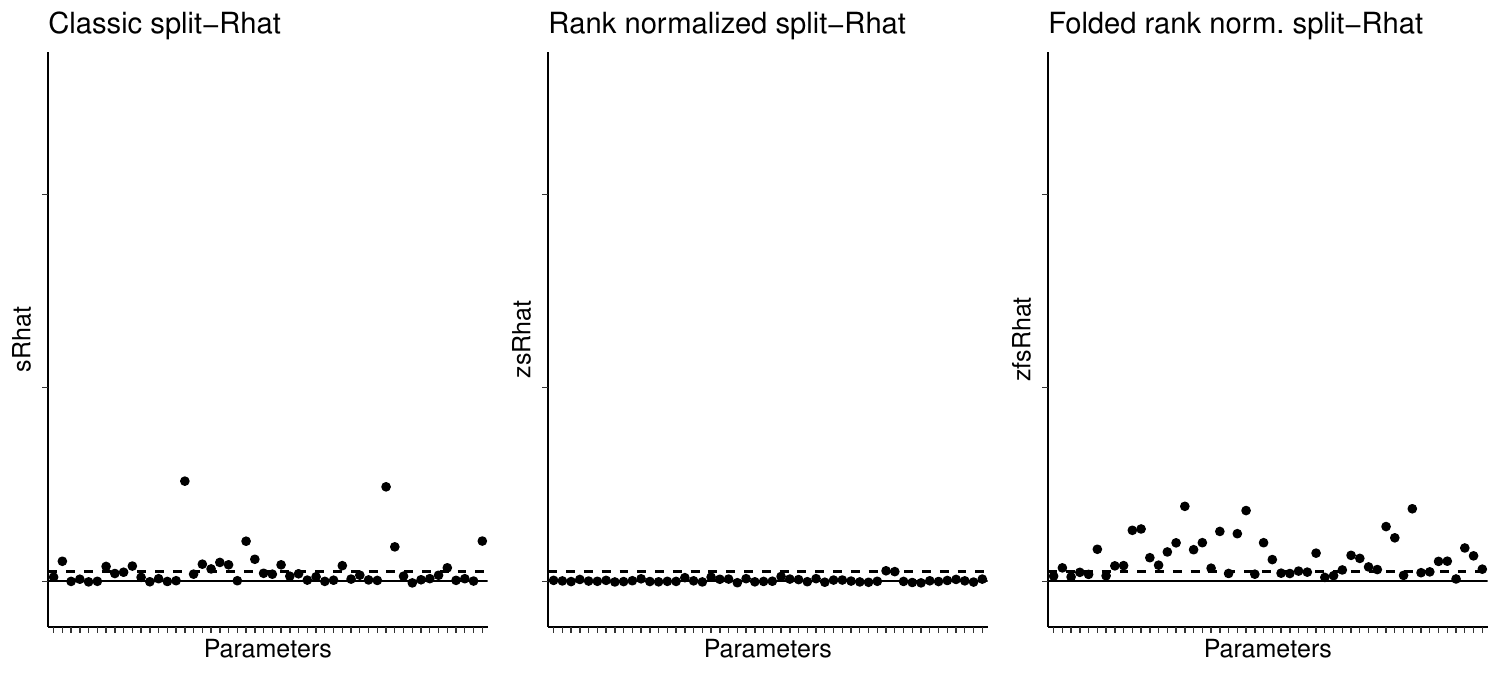}
  \caption{Traditional split-\(\widehat{R}\), rank normalized
    split-\(\widehat{R}\), and rank normalized
    folded-split-\(\widehat{R}\) for Cauchy model with nominal
    parameterization and \texttt{max\_treedepth}=20.}
  \label{fig:rhat-fit-nom-td20-1}
\end{minipage}
\hfill
  \begin{minipage}{0.48\textwidth}
  \includegraphics[width=0.98\textwidth]{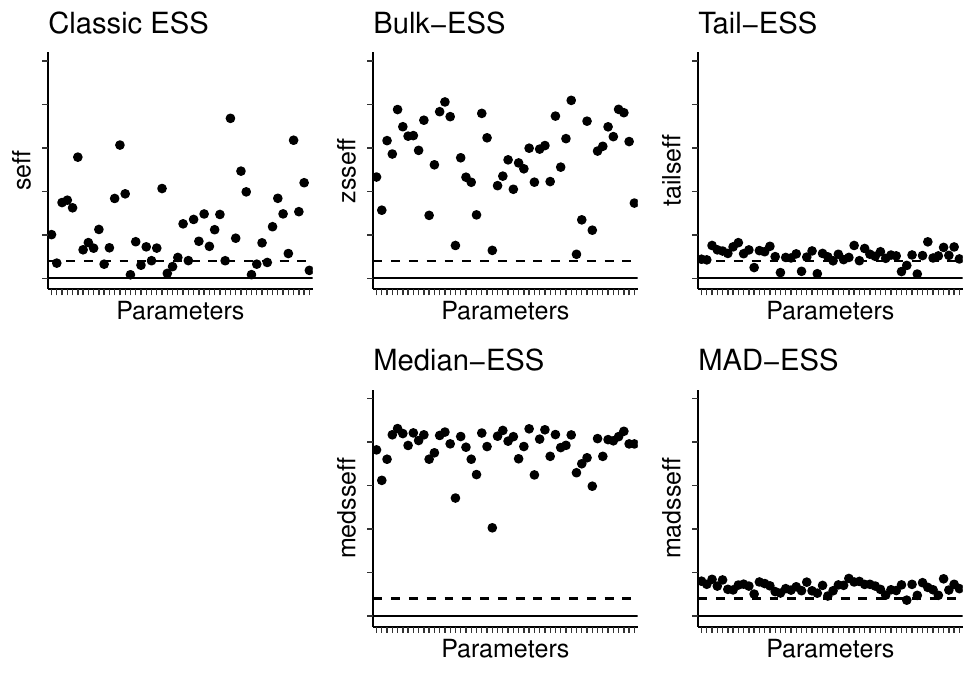}
  \caption{Traditional ESS, bulk-ESS, tail-ESS, median-ESS and MAD-ESS for
    Cauchy model with nominal parameterization \texttt{max\_treedepth}=20.\\~}
  \label{fig:ess-fit-nom-td20-1}
\end{minipage}
\end{figure}

Figure~\ref{fig:trace-fit-nom-td20-1} shows that trace plots for the first
parameter look wild with occasional large values, and it is difficult
to interpret possible convergence.
Figure~\ref{fig:rhat-fit-nom-td20-1} shows traditional \sRhat,
rank normalized \sRhat, and rank normalized
folded-split-\(\widehat{R}\) for all 50 parameters. Traditional \sRhat, which is
not well-defined in this case, has much higher variability than rank
normalized \sRhat.  Rank normalized
folded-split-\(\widehat{R}\) has higher values than rank normalized
\sRhat indicating slow mixing especially in tails.
Figure~\ref{fig:rhat-fit-nom-td20-1} shows different effective sample
size estimates for all 50 parameters. Traditional ESS, which is not well
defined in this case, has high variability. Bulk-ESS is much more
stable, and indicates that we can get reliable estimates for the
location of the posterior (except for mean). Median ESS is even more
stable with relatively high values, indicating that we can estimate
median of the distribution reliably. Tail-ESS has low values,
indicating still too slow mixing in tails for reliable tail quantile
estimates. MAD ESS values are just above our recommend threshold,
indicating practically useful MAD estimates, too. The online appendix
has additional results with longer chains, showing that all other ESS
values except traditional ESS (which is not well defined) keep improving
with more iterations. It is however recommended to use a more efficient
parameterization especially if the tail quantiles are of
interest.

\hypertarget{a-centered-eight-schools-model-1}{%
\subsection*{Appendix C: A centered eight schools model with very long chains and
thinning}\label{a-centered-eight-schools-model-1}}
\addcontentsline{toc}{subsection}{Appendix C: A centered eight schools model}

Here we demonstrate a limitation of split-\(\widehat{R}\) and ESS as
convergence diagnostics in a case where the chains seem to converge to
a common stationary distribution, but other diagnostics can detect a
likely bias.

When autocorrelation time is high, sometimes the chains are
``thinned'' by saving only a small portion of the draws.  In general
we don't recommend this approach, as it throws away useful information
leading to less efficient estimates and the dependent simulation draws
are not a problem for estimating the Monte Carlo error. However, we
also sometimes use thinning when autocorrelation time is so high that
our usual computers have memory challenges in handling unthinned
chains. This example serves as warning that thinning can also throw
away information useful for convergence diagnostics.
As in Section \ref{a-centered-eight-schools-model} we run HMC for the
eight schools model with centered parameterization, but now with
$4\times 10^5$ iterations per chain, first half removed as warm-up,
and the second half thinned by keeping only every 200th iteration.

We observe several divergent transitions and the estimated Bayesian
fraction of missing information \citep{betancourt2017conceptual} is
also low, which indicate convergence problems. In Section
\ref{a-centered-eight-schools-model} we demonstrated that the
diagnostics discussed in this paper are also able to detect
convergence problems.

\begin{figure}[tp]
  \centering
  \begin{minipage}{0.48\textwidth}
  \includegraphics[width=0.98\textwidth]{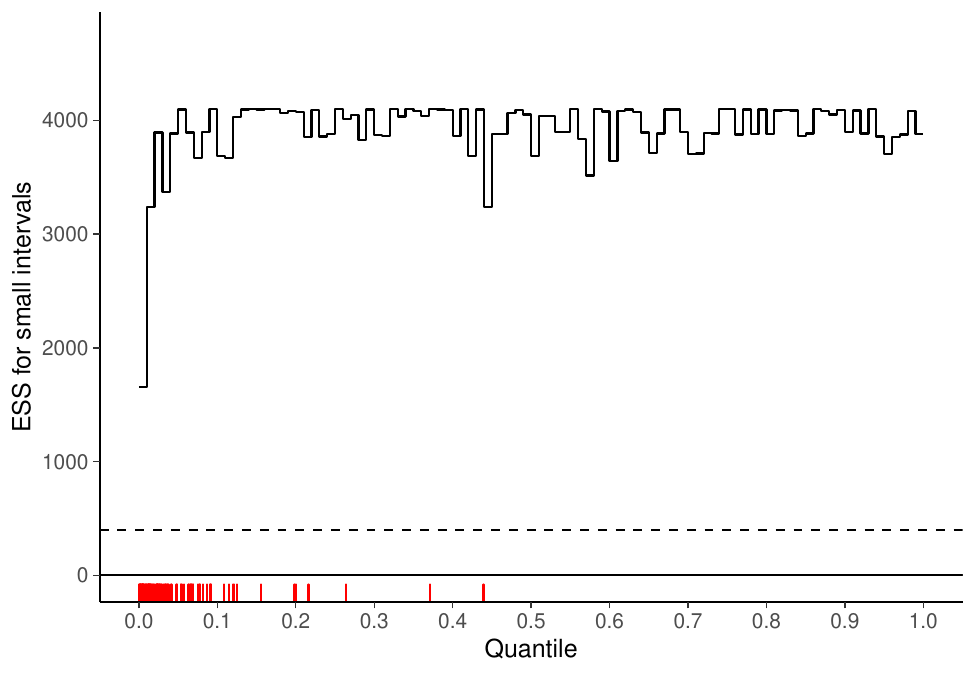}
  \caption{Local efficiency of small-interval probability estimates for eight
  schools model with centered parameterization, very long chains, and thinning. The dashed line shows the
    recommended threshold of $400$.}
  \label{fig:local-ess-fit-cp4-tau-1}
\end{minipage}
\hfill
  \begin{minipage}{0.48\textwidth}
  \includegraphics[width=0.98\textwidth]{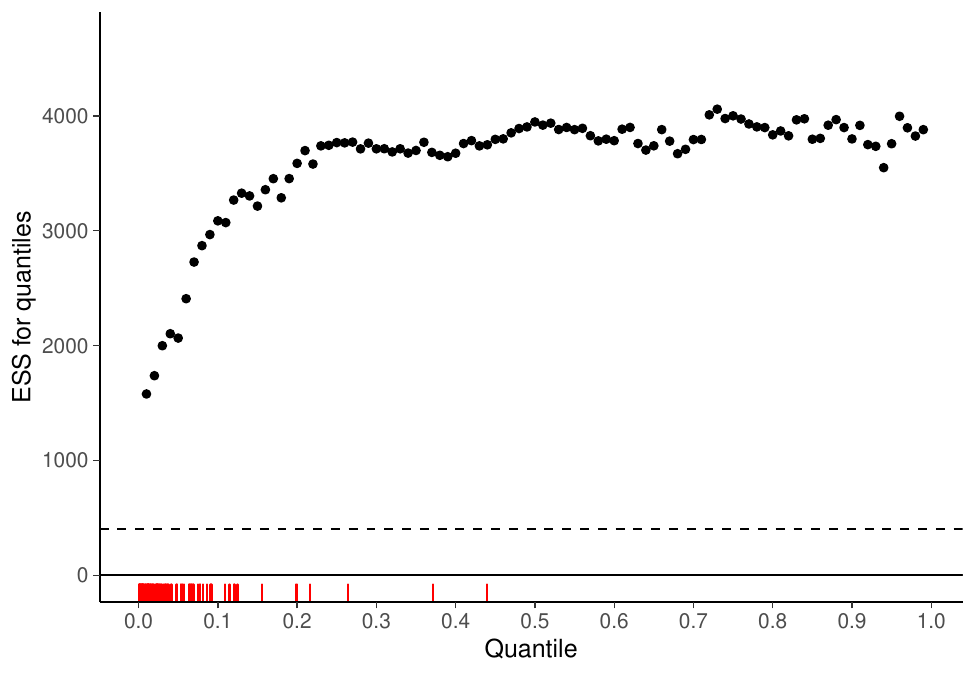}
  \caption{Efficiency of quantile estimates for eight schools model with 
  centered parameterization, very long chains, and thinning. The dashed line shows the
    recommended threshold of $400$.}
  \label{fig:quantile-ess-fit-cp4-tau-1}
\end{minipage}
\end{figure}

\begin{figure}[tp]
  \centering
  \begin{minipage}{0.48\textwidth}
  \includegraphics[width=0.98\textwidth]{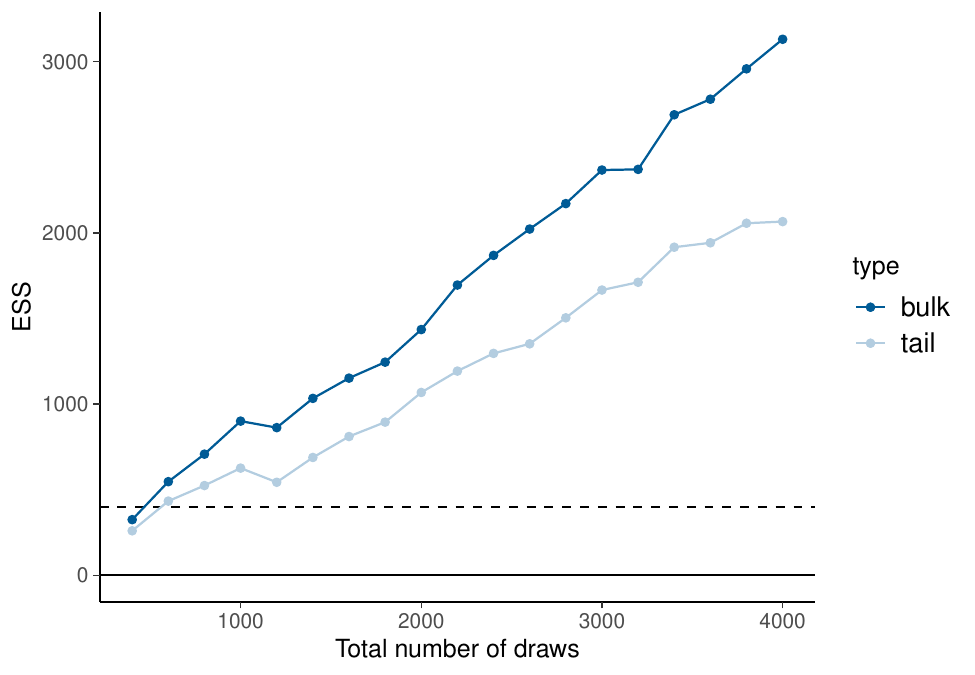}
  \caption{Estimated effective sample sizes with increasing number of 
  iterations for eight schools model with centered parameterization, very 
  long chains, and thinning. The dashed line shows the
    recommended threshold of $400$.}
  \label{fig:change-ess-fit-cp4-tau-1}
\end{minipage}
\hfill
  \begin{minipage}{0.48\textwidth}
  \includegraphics[width=0.98\textwidth]{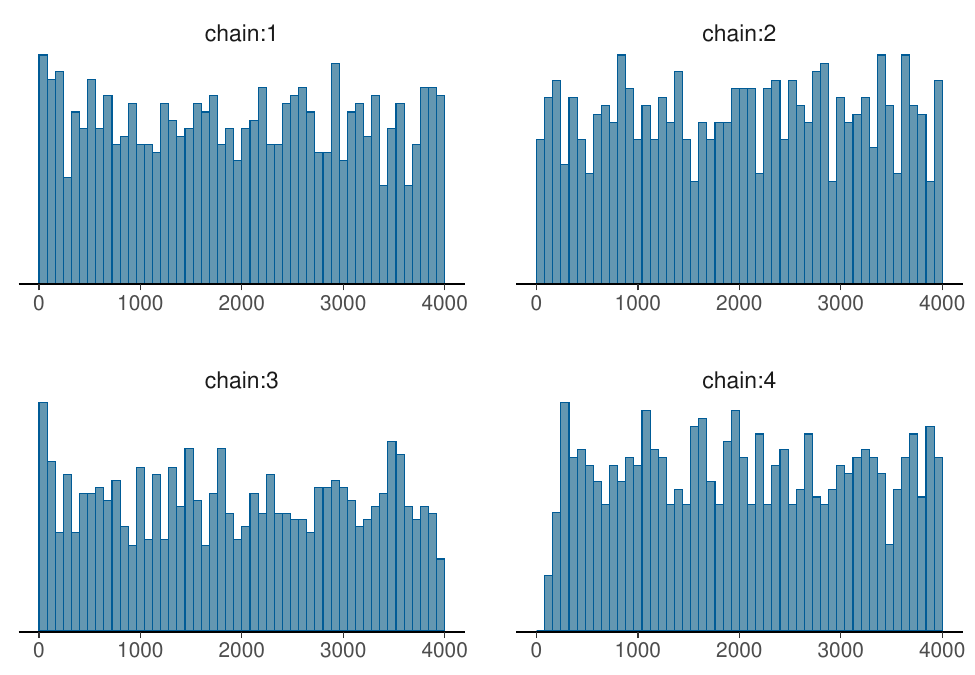}
  \caption{Rank plots of posterior draws from four chains for 8
    schools model with centered parameterization, very long chains, and
    thinning.}
  \label{fig:hist-fit-cp4-tau-1}
\end{minipage}
\end{figure}

Figures \ref{fig:local-ess-fit-cp4-tau-1},
\ref{fig:quantile-ess-fit-cp4-tau-1}, and
\ref{fig:change-ess-fit-cp4-tau-1} show the efficiency of small
probability interval estimates, efficiency of quantile estimates, and
change of bulk-ESS and tail-ESS with increasing number of iterations.
Unfortunately, after thinning, split-\(\widehat{R}\) and ESS
miss the problems. The posterior mean is still off,
being more than 3 standard deviations away from the estimate obtained using
non-centered parameterization. In this case all four chains fail
similarly in exploring the narrowest part of the funnel and all
chains seem to ``converge'' to a wrong stationary distribution.
However, the rank plots shown in Figure~\ref{fig:hist-fit-cp4-tau-1}
are still able to show the problem.

An explanation for the changed behavior after thinning is that we are
throwing away information which would make it easier to see
``sticking'' behavior in autocorrelations.  When MCMC struggles to
reach some part of the parameter space that has substantial posterior
mass, it is not unusual for Markov chains to stick for several
iterations \citep[see, e.g.][]{Neal:2003,Betancourt+Girolami:2019}. When
case sticking occurs, we usually can observe high variation in means
of chains leading to high \Rhat\ values. In infinite time, all chains
would sample from the target distribution. With long but finite
chains we can observe a situation where chains start to resemble each
other, but all are still producing biased estimates. This is clearly a
failure mode for \Rhat\ and the failure seems to be more likely when
thinning is discarding useful information about autocorrelations of
the original chains. Fortunately, we have diagnostics for HMC that
are specifically sensitive to cases where sticking tends to occur.

\hypertarget{eight-schools-with-jags}{%
\subsection*{Appendix D: A centered eight schools model fit using a Gibbs sampler}\label{eight-schools-with-jags}}
\addcontentsline{toc}{subsection}{Appendix D: Eight Schools with JAGS}

So far, we have run all models in Stan, but here we demonstrate that
these diagnostics are  also useful for samplers other than 
Hamiltonian Monte Carlo.  We fit the eight schools models also with
 JAGS \citep{plummer2003jags}, which uses a dialect of the BUGS
language \citep{BUGSproject:2009} to specify models. JAGS uses a 
mix of Gibbs and Metropolis-Hastings sampling which often does
not scale well to high-dimensional posteriors
\citep[see, e.g.][]{Hoffman+Gelman:2014} but can
work fine for relatively simple models such as in this case study.

First, we sample 1000 iterations for each of the 4 chains for easy
comparison with the corresponding Stan results. Examining the
diagnostics for $\tau$, split-\(\widehat{R}=1.08\), bulk-ESS$=59$, and
tail-ESS$=53$. 1000 iterations is clearly not enough. The online
appendix shows also the usual visual diagnostics for 1000 iterations
run, but here we report the results with 10\,000 iterations.
Examining the diagnostics for $\tau$, now split-\(\widehat{R}=1.01\),
bulk-ESS $=677$, and tail-ESS $=1027$, which are all good.

\begin{figure}[tp]
  \centering
  \begin{minipage}{0.48\textwidth}
  \includegraphics[width=0.98\textwidth]{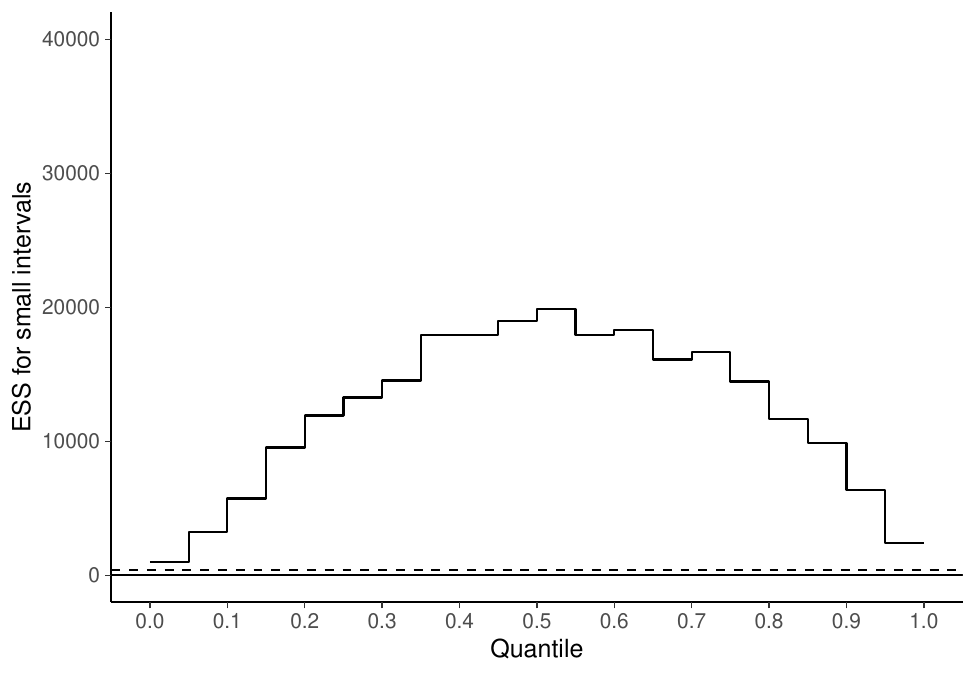}
  \caption{Local efficiency of small-interval probability estimates for the 
  eight schools model with centered parameterization and Gibbs sampling. The dashed line shows the
    recommended threshold of $400$.}
  \label{fig:local-ess-jags-cp-tau-longer-1}
\end{minipage}
\hfill
  \begin{minipage}{0.48\textwidth}
  \includegraphics[width=0.98\textwidth]{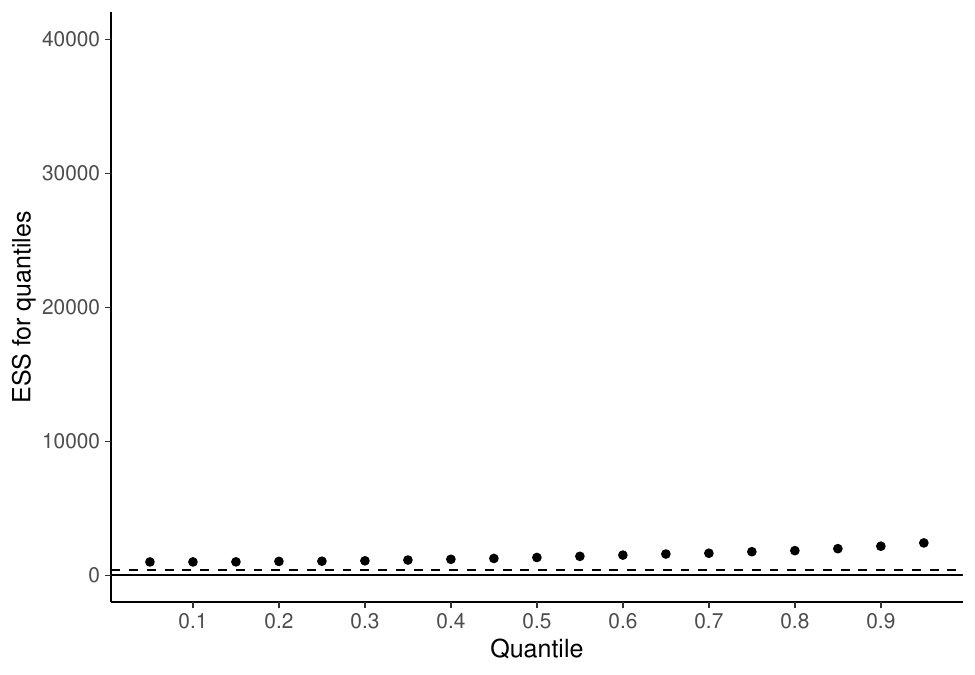}
  \caption{The efficiency of quantile estimates for the eight schools model with
  centered parameterization and Gibbs sampling. The dashed line shows the
    recommended threshold of $400$.}
  \label{fig:quantile-ess-jags-cp-tau-longer-1}
\end{minipage}
\end{figure}

\begin{figure}[tp]
  \centering
  \includegraphics[width=0.47\textwidth]{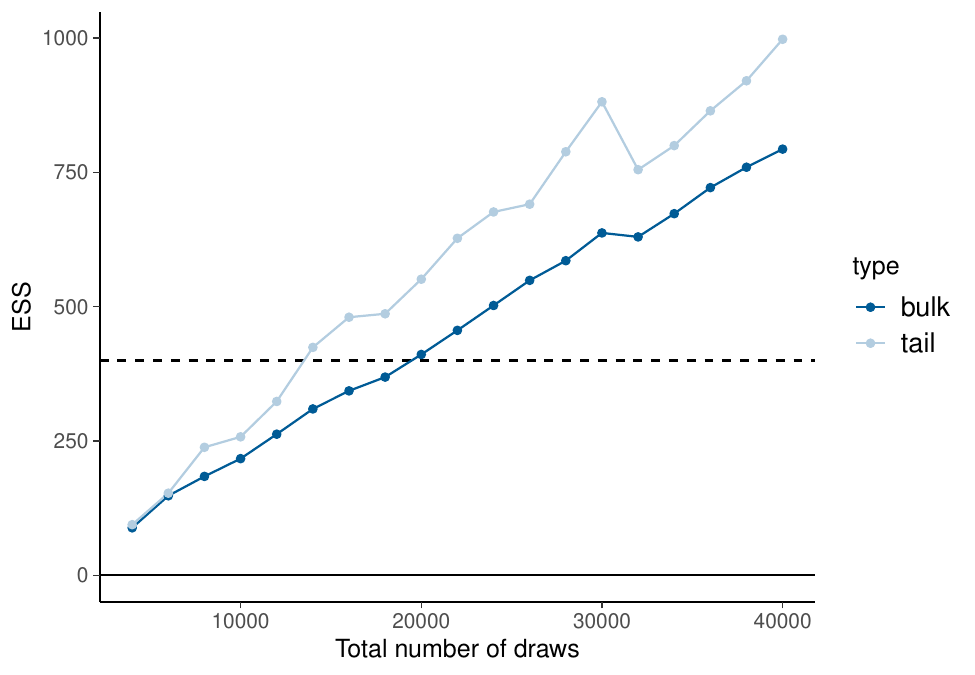}
  \caption{Change in bulk-ESS and tail-ESS with increasing number of iterations for
  the eight schools model with centered parameterization and Gibbs sampling. The dashed line shows the
    recommended threshold of $400$.}
  \label{fig:change-ess-jags-cp-tau-longer-1}
\end{figure}

Figures \ref{fig:local-ess-jags-cp-tau-longer-1},
\ref{fig:quantile-ess-jags-cp-tau-longer-1}, and
\ref{fig:change-ess-jags-cp-tau-longer-1} show the efficiency of small
probability interval estimates, efficiency of quantile estimates, and
change of bulk-SS and tail-ESS with increasing number of
iterations. The relative efficiency is low, but ESS for all small
probability intervals, quantiles and bulk are above the recommend
threshold. Notably, the increase in effective sample size for
$\tau$ is linear in the total number of draws.  A Gibbs sampler can
reach the narrow part of the funnel, although the sampling efficiency
is affected by the funnel \citep{Gelman+Huang+vanDyk+Boscardin:2008}.
In this simple case the inefficiency of the
Gibbs sampling is not dominating and good results can be achieved in
reasonable time. The online appendix shows additional results for Gibbs
sampling with a more efficient non-centered parameterization.

\end{document}